\documentclass[a4paper,12pt,parskip=half,english,numbers=noenddot]{scrartcl}

\usepackage{babel}
\usepackage[T1]{fontenc}
\usepackage[utf8]{inputenc}   % falls pdflatex genutzt wird
\usepackage{lmodern}
\usepackage{textcomp,latexsym}
\usepackage{verbatim}
\usepackage{cancel}
\usepackage{geometry}
\DeclareOldFontCommand{\it}{\normalfont\itshape}{\mathit}

\usepackage{amsmath,amssymb,amsthm}
\usepackage{bm}
\usepackage{bbm}

\usepackage{paralist}
\usepackage{enumerate}
\usepackage{threeparttable}
\usepackage{longtable,booktabs,multirow}
\usepackage{tabularx}
\usepackage{nth}
\usepackage{alphalph}

\usepackage[footnotesize,labelfont={bf}]{caption}
\usepackage[normalem]{ulem}
\usepackage{upgreek}

\usepackage{graphicx}
\usepackage{xcolor}
\graphicspath{{./pictures/}}

\usepackage{tikz}
\usetikzlibrary{
  decorations.pathmorphing,
  matrix,
  arrows,
  arrows.meta,
  calc,
  shapes,
  shapes.geometric
}

\usepgfmodule{nonlineartransformations}
\usepgflibrary{curvilinear}

\usepackage{tikz-3dplot}

\tikzstyle{block}     = [draw,rectangle,thick,minimum height=2em,minimum width=2em]
\tikzstyle{sum}       = [draw,circle,inner sep=0mm,minimum size=2mm]
\tikzstyle{connector} = [->,thick]
\tikzstyle{line}      = [thick]
\tikzstyle{branch}    = [circle,inner sep=0pt,minimum size=1mm,fill=black,draw=black]
\tikzstyle{guide}     = []
\tikzstyle{snakeline} = [
  connector,
  decorate,
  decoration={
    pre length=0.2cm,
    post length=0.2cm,
    snake,
    amplitude=.4mm,
    segment length=2mm
  },
  thick, magenta, ->
]

\usepackage{hyperref}
\hypersetup{hidelinks}
\usepackage[boxed]{algorithm2e}

\DeclareCaptionStyle{table_new}{font={footnotesize},justification=raggedright}

\setdefaultenum{1)}{\theenumi.1)}{}{}

\makeatletter
\renewcommand*{\env@matrix}[1][*\c@MaxMatrixCols c]{%
  \hskip -\arraycolsep
  \let\@ifnextchar\new@ifnextchar
  \array{#1}}
\makeatother

\newtheorem{remark}{Remark}

\usepackage{scalerel}

\renewcommand{\vec}{\boldsymbol}   % <-- konsistent

\newcommand{\ten}[1]{\pmb{#1}}

\renewcommand{\d}[1][]{\,\mathrm{d}#1}

\newcommand{\F}{\mathcal{F}}

\newcommand{\R}{\mathbb{R}}

\newcommand{\SO}{\operatorname{SO}}

\newcommand{\K}{\mathcal{K}}

\newcommand{\axl}{\operatorname{axl}}

\newcommand{\e}{^{\scriptstyle \mathrm{e}}}

\newcommand{\s}{\mathfrak{s}}

\newcommand{\tp}{^{\scriptstyle \mathrm{T}}}

\newcommand{\VPTen}[1]{\left[#1\right]_{\times}}
\renewcommand{\det}{\operatorname{det}}

\newcommand{\mi}[1]{\vec{#1}}

\renewcommand{\ten}[1]{\bm{#1}}

\usepackage[section]{placeins}

\usepackage{lineno}
\begin{document}

\begin{center}
  {\Large\bfseries Space--time formulation of geometrically exact beams\par}
  \vspace{1.0em}
  {C. Hesch$^{\ast}$, E. Sahin}\\
  \vspace{0.4em}
  {Chair of Computational Mechanics, University of Siegen, Siegen, Germany\par}
\end{center}

\begin{abstract}
We present a configuration–momentum space–time formulation for geometrically exact Cosserat beams with curved and twisted stress-free reference configurations. The beam history is represented as a directed world sheet in non-relativistic \(\mathbb R^4\), with prescribed absolute time. Spatial resultants and temporal momenta form common fluxes whose co-normal boundary operator unifies spatial end loads, temporal inflow and outflow while leaving the configuration free at the prescribed final time. Starting from a continuous mixed action, we discretize the rearranged first-order equations using continuous time-aligned tensor-product elements, future-directed Petrov–Galerkin tests and multiplicative rotation updates on \(\mathrm{SO}(3)\).

Numerical studies assess temporal-boundary consistency, convergence and covariance under constant superposed rotations, complemented by modal tests and implicit-midpoint comparisons. A helical spring coupled to a rigid body combines short-time refinement with large motions over two seconds. Future-directed stabilization reduces the response before delayed load activation, without establishing strict causality. The investigated continuous \(P_2/P_2\) simplex construction exhibits strong sensitivity to mesh resolution and the chosen triangulation pattern. These findings support the time-aligned construction and identify stable spaces on non-aligned meshes as an open discretization issue.
\end{abstract}

\noindent\textbf{Keywords:}
space--time FEM; geometrically exact beam; Cosserat beam;
future-directed stabilization; co-normal flux; frame covariance

\noindent$^{\ast}$Corresponding author. E-mail address: christian.hesch@uni-siegen.de

\section{Introduction}
\label{sec:introduction}

Geometrically exact beam and rod models describe slender structures
undergoing finite displacements and large rotations by a centerline and
an orthonormal director frame. They retain extension, shear, bending and
torsion without linearizing the rotational kinematics, providing a
mechanical description of flexible structural members and curved
components such as springs. The classical formulations and their
continuum foundations are well established
\cite{reissner_finite_1981,simo_finite_1985,simo_three-dimensional_1986,antman_nonlinear_2005,auricchio_geometrically_2008}.
Their numerical realization requires compatible treatments of spatial
rotation interpolation, temporal evolution and boundary data.
Objective interpolation and structure-preserving integration have
therefore received sustained attention
\cite{romero_objective_2002,romero_interpolation_2004,betsch_frameindifferent_2002,armero_energy-dissipative_2003,leyendecker_objective_2006,eugster_directorbased_2014}.
Lie-group methods provide a general setting for preserving the rotation
manifold under updates \cite{iserles_lie-group_2000}. Recent developments
include spherical B\'ezier interpolation for time integration
\cite{cammarata_implicit_2025}, mixed elements with independent moment
resultants and discontinuous rotations \cite{humer_mixed_2026}, and
port-Hamiltonian formulations of Cosserat-rod dynamics
\cite{kinon_mixed_2026}.

Simultaneous finite-element approximations in space and time have been
developed for fluid mechanics and moving domains
\cite{shakib_new_1991-1,tezduyar_spacetime_2006,tezduyar_spacetime_2010,tezduyar_spacetime_2019},
elastodynamics and variational mechanics
\cite{hughes_space-time_1988,borri_general_1993,abedi_spacetime_2006,idesman_solution_2007,hesch_variational_2017},
and parabolic evolution problems \cite{andreev_stability_2012,steinbach_space-time_2015}.
The present work continues the space--time mechanics framework for
nonlinear continua \cite{schus_nonlinear_2023} and rigid multibody
dynamics \cite{hesch_space-time_2024}. A closely related beam formulation
is due to Steinbrecher and Humer \cite{steinbrecher_space-time_2026}.
Their mixed approach uses independent translational and material angular
velocities, objective rotation interpolation, and supports curved beams,
rigid joints and time-dependent material domains.
Our contribution concerns a continuous configuration--momentum action
and a common boundary representation of spatial resultants and temporal
momenta, including the outgoing momentum at a free terminal
configuration.

The beam history is represented by the two-dimensional surface
\[
  \Xi(s,t)=\bigl(\mi r(s,t),c_\tau t\bigr)\in\mathbb R^4,
\]
termed a directed world sheet. Time retains its classical,
non-relativistic meaning: \(t\) is prescribed and is held fixed under
mechanical variations. The constant \(c_\tau>0\) converts time to length
units; it is not a material wave speed, and the fourth coordinate adds
no mechanical degree of freedom. The fixed reference beam may be curved
and initially twisted. Its director frame \(\ten\Lambda_0(s)\) enters
through the relative rotation \(\ten R=\ten\Lambda\ten\Lambda_0\tp\),
retaining the reference geometry explicitly in the strain measures.

On this surface, the force resultant \(\mi n\) and linear momentum
\(\mi p\) form the spatial and temporal parts of a common force flux;
the couple resultant \(\mi m\) and intrinsic angular momentum
\(\mi\ell\) form the corresponding moment flux. The boundary co-normal
is tangent to the surface and normal to its boundary curve. Contracting
the fluxes with this co-normal expresses spatial end loads and temporal
momentum transfer in one operator. At a prescribed final time, the
terminal configuration remains free and the outgoing flux is identified
with its interior trace. This completion avoids the artificial
zero-momentum condition that would follow from free terminal variations
without an outgoing flux.

The temporal momenta are independent fields in the continuous mixed
action. Its rotational variation couples the configuration equation to
the angular momentum--velocity compatibility equation. A continuous
rearrangement removes this coupling without changing the continuous
solution set, and the resulting first-order system is discretized.
Since this rearrangement and finite-dimensional projection need not
commute, momentum--velocity compatibility and mechanical balance defects
are monitored separately from the algebraic residual.

The principal realization uses continuous tensor-product spaces aligned
with the time foliation, with \(Q_2/Q_2\) as the quadratic baseline and
multiplicative rotation updates on \(\SO(3)\). A globally coupled
approximation can respond before a delayed load is activated.
We perturb all four test fields in the prescribed future direction,
using one scalar parameter and no additional interior-facet parameter
for the aligned construction. This choice is related to established
Petrov--Galerkin stabilization: SUPG biases tests along a transport
direction \cite{brooks_streamline_1982}, whereas Galerkin/least-squares
methods add residual-based least-squares terms \cite{hughes_new_1989}.
Discontinuous space--time methods provide an alternative with element
interfaces and numerical fluxes \cite{abedi_spacetime_2006}. Here the
perturbation follows the absolute-time foliation and is assessed through
the pre-response--accuracy trade-off. Neither strict causality nor the
stability properties of those other methods are inferred from this
connection.

The numerical studies separate temporal outflow, smooth-solution
convergence, spatial shear effects, temporal accuracy and covariance
under constant superposed rotations. A curved helical spring coupled to
a rigid body tests the formulation beyond straight reference beams,
using short-time refinement and an implicit-midpoint comparison; a
longer trajectory illustrates the motion. A non-aligned-facet extension
has pairwise conservative numerical fluxes under the stated assumptions,
but the selected continuous equal-order \(P_2/P_2\) simplex construction
exhibits strong sensitivity to mesh resolution and the chosen
triangulation pattern. These observations support the aligned
construction as the numerical baseline studied here.

Section~\ref{sec:cosserat_dictionary} fixes the beam notation and reference
geometry. Section~\ref{sec:world_sheet} develops the world sheet, common
fluxes and continuous mixed formulation. Section~\ref{sec:spacetime_discretization}
presents the aligned discretization and its diagnostics,
Section~\ref{sec:numerical_examples} reports the numerical studies, and
Section~\ref{sec:conclusions} gives the conclusions.
Appendix~\ref{app:stabilization_details} provides the temporal completion,
non-aligned facets and alternative momentum spaces.

\section{Cosserat beam data: a minimal three-dimensional dictionary}
\label{sec:cosserat_dictionary}

This section collects the classical Cosserat-beam kinematics and constitutive quantities used in the subsequent \(\R^4\) formulation \cite{reissner_finite_1981,simo_finite_1985,simo_three-dimensional_1986}.
We specify the admissible configurations and virtual fields, the initial-boundary data, and the integral balance statement that connects the classical beam problem to the space--time flux formulation of Section~\ref{sec:world_sheet}.

Bold lowercase symbols denote spatial vectors; bold capitals denote second-order tensors or explicitly defined reference and material vectors, such as \(\mi X\), \(\mi D_i\), \(\mi\Gamma\), \(\mi N\), and \(\mi M\).
The calligraphic symbol \(\mathcal K\) used below denotes an elastic curvature-change vector, distinct from the scalar kinetic-energy density \(K\).
A comma followed by \(s\) or \(t\) denotes a partial derivative at fixed \(t\) or \(s\), respectively.
Repeated director indices \(i,j\in\{1,2,3\}\) and cross-section indices \(\alpha,\beta\in\{1,2\}\) are summed unless otherwise stated.
For \(\mi a\in\R^3\), the skew tensor \(\VPTen{\mi a}\) is defined by
\[
  \VPTen{\mi a}\,\mi b=\mi a\times\mi b
  \qquad \forall\,\mi b\in\R^3,
\]
and \(\axl(\VPTen{\mi a})=\mi a\).
The rotation group is
\[
  \SO{}(3)=\{\ten R\in\R^{3\times3}\,|\,\ten R\tp\ten R=\ten I,\;\det\ten R=1\}.
\]

\subsection{Configuration and directors}
\label{subsec:dictionary_configuration}

Let
\[
  \mathfrak I := (0,L),
  \qquad
  \mathcal T := (0,t_{\mathrm{end}}),
  \qquad
  \mathcal B_0 := \mathfrak I\times\mathcal T,
\]
where \(L>0\) is the reference beam length, \(t_{\mathrm{end}}>0\) the final time, \(s\) the reference arc-length coordinate, and \(t\) absolute time.
As in classical beam dynamics, \(t\) is an independent parameter held fixed under mechanical variations; no equation is solved for time itself.
The embedding in Section~\ref{subsec:worldsheet_embedding} records this same parameter as a fourth coordinate without adding a mechanical degree of freedom.
The smooth, time-independent reference centerline is denoted by
\[
  \mi r_0:\mathfrak I\rightarrow\R^3,
  \qquad
  \|\mi r_{0,s}\|=1,
\]
and is equipped with a smooth, positively oriented orthonormal reference director triad
\[
  \{\mi D_1(s),\mi D_2(s),\mi D_3(s)\}.
\]
The reference is unstretched and unsheared, with
\[
  \mi r_{0,s}=\mi D_3(s),
\]
but may be curved and initially twisted; the reference directors need not be constant in \(s\).
All reference data are held fixed under time evolution and mechanical variations.
The reference offset vector \(\mi\Theta\) and the full reference position \(\mi X\) are
\[
  \mi\Theta(s,\Theta^\alpha)
  :=\Theta^\alpha\mi D_\alpha(s),
  \qquad
  \mi X(s,\Theta^\alpha)=\mi r_0(s)+\mi\Theta(s,\Theta^\alpha).
\]
The current configuration is represented by the centerline
\[
  \mi r :=\mi r(s,t)
\]
and a rotation field
\[
  \ten R(s,t)=\mi d_i(s,t)\otimes\mi D_i(s)\in\SO{}(3),
  \qquad
  \mi d_i(s,t)=\ten R(s,t)\mi D_i(s).
\]
Thus \(\ten R\) is a relative rotation; the reference state has \(\mi r=\mi r_0\) and \(\ten R=\ten I\), even for a curved beam.
For a fixed positively oriented orthonormal basis \(\{\mi e_i\}\), define the auxiliary absolute triads
\[
  \ten\Lambda_0:=\mi D_i\otimes\mi e_i,
  \qquad
  \ten\Lambda:=\mi d_i\otimes\mi e_i=\ten R\ten\Lambda_0.
\]
Only \(\ten R=\ten\Lambda\ten\Lambda_0\tp\) is an independent rotation field.
The stress-free reference need not coincide with the prescribed initial configuration, which may be predeformed.
The geometrically exact Cosserat ansatz for a material point then reads
\begin{equation}
  \mi x(s,\Theta^\alpha,t)
  = \mi r(s,t)+\ten R(s,t)\mi\Theta(s,\Theta^\alpha).
  \label{eq:dictionary_beam_kinematics}
\end{equation}
The current directors are independent of the current centerline tangent; hence the model is of Cosserat, or Simo--Reissner, type and includes shear deformation.
Kirchhoff-type beams are obtained only after imposing an additional shear-free constraint.

\subsection{Admissible configurations, boundary partitions, and virtual fields}
\label{subsec:dictionary_spaces_variations}

At the beam ends, \(\Gamma_r^D\) and \(\Gamma_R^D\) denote the subsets with prescribed centerline position and cross-section orientation, respectively.
Their complements \(\Gamma_r^N\) and \(\Gamma_R^N\) carry prescribed end forces and end couples.
Each Dirichlet--Neumann partition is disjoint, but the translational and rotational partitions need not coincide:
\begin{equation}
  \partial\mathfrak I
  =\Gamma_r^D\mathbin{\cup}\Gamma_r^N
  =\Gamma_R^D\mathbin{\cup}\Gamma_R^N,
  \qquad
  \nu_s(0)=-1,
  \quad
  \nu_s(L)=1,
  \label{eq:dictionary_boundary_partition}
\end{equation}
Here \(\nu_s\) is the outward normal sign of the reference interval.
For a beam clamped at \(s=0\) and free at \(s=L\), both Dirichlet subsets are \(\{0\}\) and both Neumann subsets are \(\{L\}\), with zero prescribed end force and couple at the free end.
The associated lateral parts of the parameter-domain boundary are
\[
  \Sigma_r^D:=\Gamma_r^D\times\mathcal T,
  \qquad
  \Sigma_r^N:=\Gamma_r^N\times\mathcal T,
  \qquad
  \Sigma_R^D:=\Gamma_R^D\times\mathcal T,
  \qquad
  \Sigma_R^N:=\Gamma_R^N\times\mathcal T,
\]
and the initial and terminal faces are denoted by
\[
  \Sigma_0:=\mathfrak I\times\{0\},
  \qquad
  \Sigma_T:=\mathfrak I\times\{t_{\mathrm{end}}\}.
\]
Figure~\ref{fig:dictionary_parameter_boundary} illustrates these boundary sets for a clamped--free beam.

\begin{figure}[tbp]
  \centering
  \begin{tikzpicture}[x=0.85cm,y=0.85cm,>=Stealth,font=\small]
    \fill[black!3] (0,0) rectangle (7,3.8);
    \draw[very thick] (0,0) -- (0,3.8);
    \draw[thick] (7,0) -- (7,3.8);
    \draw[thick] (0,0) -- (7,0);
    \draw[thick,dashed] (0,3.8) -- (7,3.8);
    \draw[->] (0,-0.5) -- (7.9,-0.5) node[right] {\(s\)};
    \draw[->] (-0.5,0) -- (-0.5,4.6) node[above] {\(t\)};
    \draw (0,-0.43) -- (0,-0.57) node[below] {\(0\)};
    \draw (7,-0.43) -- (7,-0.57) node[below] {\(L\)};
    \draw (-0.43,3.8) -- (-0.57,3.8)
      node[left] {\(t_{\mathrm{end}}\)};
    \node[align=center] at (3.5,2.8)
      {\(\mathcal B_0=\mathfrak I\times\mathcal T\)};
    \draw[->,thick] (3.5,0.9) -- (3.5,2.1)
      node[midway,right] {future};
    \node[anchor=east,align=center,text width=2.8cm] at (-0.9,1.9)
      {\(\Sigma_r^D=\Sigma_R^D\)\\prescribed position\\and orientation};
    \node[anchor=west,align=center,text width=2.7cm] at (7.3,1.9)
      {\(\Sigma_r^N=\Sigma_R^N\)\\zero end force\\and couple};
    \node[anchor=south,align=center] at (3.5,4.0)
      {\(\Sigma_T\): fixed final time\\free terminal configuration};
    \node[anchor=north,align=center] at (3.5,-1.1)
      {\(\Sigma_0\): prescribed initial configuration and momenta\\
       \(\mi\eta=\mi0,\quad\mi\theta=\mi0\)};
  \end{tikzpicture}
  \caption{Material space--time domain and boundary sets for the illustrative
    clamped--free case. The translational and rotational lateral partitions
    coincide in this example; they are independent in the general formulation.
    The final time is prescribed, while the terminal configuration and outgoing
    momenta are determined by the solution.}
  \label{fig:dictionary_parameter_boundary}
\end{figure}

Boundary equalities below are understood in the trace sense.
The Sobolev classes specify the kinematic boundary constraints; the geometric differential identities and classical localization arguments are derived for sufficiently smooth configurations, resultants, and data.
The additional nondegeneracy assumption for the world-sheet metric is stated in Subsection~\ref{subsec:worldsheet_embedding}.

For later use, we set
\[
  H^1(\mathcal B_0;\SO{}(3))
  :=\left\{
      \ten R\in H^1(\mathcal B_0;\R^{3\times3})
      \;\middle|\;
      \ten R(s,t)\in\SO{}(3)\ \text{a.e. in }\mathcal B_0
    \right\}.
\]
Given translational and rotational Dirichlet data \(\bar{\mi r}\), \(\bar{\ten R}\), and initial configurations \(\mi r^{\mathrm{in}}\), \(\ten R^{\mathrm{in}}\), the admissible configuration spaces are
\begin{align}
  \mathcal Q_r
  :=\bigl\{
      \mi r\in H^1(\mathcal B_0;\R^3)
      \;\big|\;
      \mi r=\bar{\mi r}\ \text{on }\Sigma_r^D,
      \ \mi r=\mi r^{\mathrm{in}}\ \text{on }\Sigma_0
    \bigr\},
  \label{eq:dictionary_solution_space_r}
  \\
  \mathcal Q_R
  :=\bigl\{
      \ten R\in H^1(\mathcal B_0;\SO{}(3))
      \;\big|\;
      \ten R=\bar{\ten R}\ \text{on }\Sigma_R^D,
      \ \ten R=\ten R^{\mathrm{in}}\ \text{on }\Sigma_0
    \bigr\}.
  \label{eq:dictionary_solution_space_R}
\end{align}
The prescribed data are assumed to satisfy the usual compatibility conditions at the intersections of the spatial and initial Dirichlet boundaries.
For smooth data, this includes compatibility of the initial velocities with the time derivatives of any prescribed boundary motion.
The virtual centerline displacement \(\mi\eta\) and spatial infinitesimal rotation (virtual spin) \(\mi\theta\) belong to the linear spaces
\begin{align}
  \mathcal V_r
  &:=\bigl\{
      \mi\eta\in H^1(\mathcal B_0;\R^3)
      \;\big|\;
      \mi\eta=\mi0\ \text{on }\Sigma_r^D\cup\Sigma_0
    \bigr\},
  \label{eq:dictionary_test_space_r}
  \\
  \mathcal V_\theta
  &:=\bigl\{
      \mi\theta\in H^1(\mathcal B_0;\R^3)\cap L^\infty(\mathcal B_0;\R^3)
      \;\big|\;
      \mi\theta=\mi0\ \text{on }\Sigma_R^D\cup\Sigma_0
    \bigr\}.
  \label{eq:dictionary_test_space_theta}
\end{align}
No condition is imposed on these fields at \(\Sigma_T\). For the initial-value problem the terminal configuration is not prescribed; the corresponding outgoing temporal momentum fluxes will be identified in Section~\ref{sec:world_sheet}. Admissible variations are represented in spatial form by
\begin{equation}
  \delta\mi r=\mi\eta,
  \qquad
  \delta\ten R=\VPTen{\mi\theta}\ten R,
  \qquad
  (\mi\eta,\mi\theta)\in\mathcal V_r\times\mathcal V_\theta.
  \label{eq:dictionary_variations}
\end{equation}
Here \(\mi\theta\) parameterises a variation at the current \(\ten R\), not a finite rotation representation of \(\ten R\) itself.
The boundedness of the virtual spin ensures that \(\VPTen{\mi\theta}\ten R\in H^1(\mathcal B_0;\R^{3\times3})\) for \(\ten R\in\mathcal Q_R\).
Consequently,
\[
  \delta\ten R\tp=-\ten R\tp\VPTen{\mi\theta}.
\]

\subsection{Mass resultants and temporal momenta}
\label{subsec:dictionary_momenta}

Let \(A_0(s)\) denote the reference cross-sectional domain and \(\rho_0(s,\Theta^\alpha)\) the effective reference density associated with the adopted rod mass measure \(\d m=\rho_0\,\d A\,\d s\).
The line mass \(A(s)\) and the reference rotational inertia per unit length are given by
\begin{align}
  A(s) &:= \int_{A_0(s)}\rho_0\,\d A, \\
  \ten J(s) &:= \int_{A_0(s)}\rho_0\,
      \VPTen{\mi\Theta}\tp\VPTen{\mi\Theta}\,\d A.
\end{align}
We assume that \(A(s)\) and the symmetric tensor \(\ten J(s)\) are bounded and uniformly positive; hence, the inverse mass and inertia used in the temporal Legendre transform are well defined and bounded.
We retain the centroidal rod assumption: the first mass moment vanishes with respect to this measure,
\[
  \int_{A_0(s)}\rho_0\,\mi\Theta\,\d A=\mi0.
\]
Otherwise, the translational and rotational kinetic energies contain the standard coupling terms; these can be added without altering the space--time balance structure.

The reference inertia is \(\ten J(s)=\ten\Lambda_0(s)\ten J^{\mathrm{loc}}(s)\ten\Lambda_0(s)\tp\), where \(\ten J^{\mathrm{loc}}\) contains its components in the reference director frame.
Constant local section properties need not give a constant matrix in a fixed Cartesian basis.
Since \(\ten\Lambda_0\) is time-independent, the spatial angular velocity is defined by
\begin{equation}
  \VPTen{\mi\omega}:=\ten R_{,t}\ten R\tp,
  \qquad
  \mi\omega=\axl(\ten R_{,t}\ten R\tp).
  \label{eq:dictionary_omega}
\end{equation}
Its variation follows directly from \eqref{eq:dictionary_variations} as
\begin{equation}
  \delta\mi\omega
  =\mi\theta_{,t}-\mi\omega\times\mi\theta.
  \label{eq:dictionary_omega_variation}
\end{equation}
For vanishing first mass moments the kinetic energy per unit reference length is
\begin{equation}
  K
  =\frac12 A\,\mi r_{,t}\cdot\mi r_{,t}
   +\frac12 \mi\omega\cdot\ten J^{\mathrm{sp}}\mi\omega,
  \qquad
  \ten J^{\mathrm{sp}}:=\ten R\ten J\ten R\tp .
  \label{eq:dictionary_kinetic_energy}
\end{equation}
The conjugate temporal momenta are therefore
\begin{equation}
  \mi p:=\frac{\partial K}{\partial\mi r_{,t}}=A\mi r_{,t},
  \qquad
  \mi\ell:=\frac{\partial K}{\partial\mi\omega}=\ten J^{\mathrm{sp}}\mi\omega.
  \label{eq:dictionary_temporal_momenta}
\end{equation}
Both quantities are expressed in the spatial frame.
For a beam segment of reference length \(\d s\),
\(\mi p\,\d s\) and \(\mi\ell\,\d s\) are its linear and
intrinsic angular momenta, respectively.
Here, \(\mi\ell\) denotes the angular momentum of cross-section rotation about its mass center, excluding the orbital contribution \(\mi r\times\mi p\).
The total angular momentum density about a fixed spatial origin is therefore \(\mi r\times\mi p+\mi\ell\). The variation of the spatial inertia is
\begin{equation}
  \delta\ten J^{\mathrm{sp}}
  =\VPTen{\mi\theta}\ten J^{\mathrm{sp}}
   -\ten J^{\mathrm{sp}}\VPTen{\mi\theta}.
  \label{eq:dictionary_spatial_inertia_variation}
\end{equation}
Its contribution cancels the convective part of \eqref{eq:dictionary_omega_variation}, so the complete kinetic-energy variation is
\begin{equation}
  \delta K
  =\mi p\cdot\mi\eta_{,t}
   +\mi\ell\cdot\mi\theta_{,t}.
  \label{eq:dictionary_kinetic_variation}
\end{equation}
Thus \(\mi p\) and \(\mi\ell\) are the temporal fluxes conjugate to \(\mi\eta\) and \(\mi\theta\), respectively.

\subsection{Elastic strain measures, variations, and stress resultants}
\label{subsec:dictionary_strains}

For the generally curved reference configuration, the material strain measures are
\begin{equation}
  \mi\Gamma := \ten R\tp\mi r_{,s}-\mi D_3,
  \qquad
  \mathcal K := \axl(\ten R\tp\ten R_{,s}).
  \label{eq:dictionary_strains}
\end{equation}
These are referential vectors; their components along \(\mi D_i(s)\) measure axial and shear strains, and changes in bending and twist, respectively.
Differentiating \(\ten R=\ten\Lambda\ten\Lambda_0\tp\) gives
\[
  \mathcal K
  =\ten\Lambda_0\left[
      \axl(\ten\Lambda\tp\ten\Lambda_{,s})
      -\axl(\ten\Lambda_0\tp\ten\Lambda_{0,s})
    \right].
\]
Thus \(\mathcal K\) already includes the reference-curvature subtraction; no second reference term is subtracted from \(\axl(\ten R\tp\ten R_{,s})\).
Both strains vanish at \((\mi r,\ten R)=(\mi r_0,\ten I)\).
Because the reference fields are fixed under \eqref{eq:dictionary_variations}, the material variations retain the form
\begin{equation}
  \delta\mi\Gamma
  =\ten R\tp\left(\mi\eta_{,s}-\mi\theta\times\mi r_{,s}\right),
  \qquad
  \delta\mathcal K
  =\ten R\tp\mi\theta_{,s}.
  \label{eq:dictionary_material_strain_variations}
\end{equation}
The stored energy per unit reference length is written as
\begin{equation}
  \Psi=\Psi(s,\mi\Gamma,\mathcal K),
  \label{eq:dictionary_energy}
\end{equation}
and the material stress resultants and their spatial counterparts are
\begin{equation}
  \mi N:=\frac{\partial\Psi}{\partial\mi\Gamma},
  \qquad
  \mi M:=\frac{\partial\Psi}{\partial\mathcal K},
  \qquad
  \mi n:=\ten R\mi N,
  \qquad
  \mi m:=\ten R\mi M.
  \label{eq:dictionary_stress_resultants}
\end{equation}
The derivatives are taken at fixed \(s\).
At zero strain, we require \(\mi N=\mi M=\mi0\) at every \(s\) and normalize \(\Psi=0\), so that the reference is stress-free.
The material orientation is included through
\[
  \Psi(s,\mi\Gamma,\mathcal K)
  =\Psi_{\mathrm{loc}}\bigl(s,\ten\Lambda_0\tp\mi\Gamma,\ten\Lambda_0\tp\mathcal K\bigr),
\]
where \(\Psi_{\mathrm{loc}}\) is the section law in local director components.
For a quadratic law this amounts to rotating each local stiffness tensor into the reference orientation, just as for \(\ten J\).
Thus \(\mi n\) is the spatial force resultant and \(\mi m\) the spatial couple resultant transmitted through a beam cross-section.
The corresponding spatial strain vectors are denoted by
\[
  \mi\gamma:=\ten R\mi\Gamma=\mi r_{,s}-\mi d_3,
  \qquad
  \mi\kappa:=\ten R\mathcal K=\axl(\ten R_{,s}\ten R\tp).
\]
Here \(\mi\kappa\) is the spatially represented elastic curvature change, not the absolute curvature of the current director triad.
The spatial stress resultants are energetically conjugate to the pushed-forward material variations
\begin{equation}
  \delta^\sharp\mi\gamma
  :=\ten R\,\delta\mi\Gamma
  =\mi\eta_{,s}-\mi\theta\times\mi r_{,s},
  \qquad
  \delta^\sharp\mi\kappa
  :=\ten R\,\delta\mathcal K
  =\mi\theta_{,s}.
  \label{eq:dictionary_pushed_variations}
\end{equation}
The superscript \(\sharp\) distinguishes these quantities from the full variations of the spatial strain vectors, which satisfy
\begin{equation}
  \delta\mi\gamma
  =\delta^\sharp\mi\gamma+\mi\theta\times\mi\gamma,
  \qquad
  \delta\mi\kappa
  =\delta^\sharp\mi\kappa+\mi\theta\times\mi\kappa.
  \label{eq:dictionary_full_spatial_strain_variations}
\end{equation}
At a fixed time, the internal virtual work therefore reads
\begin{equation}
  \delta U^{\mathrm{int}}(t)
  =\int_{\mathfrak I}
      \left(
        \mi N\cdot\delta\mi\Gamma
        +\mi M\cdot\delta\mathcal K
      \right)\d s
  =\int_{\mathfrak I}
      \left(
        \mi n\cdot\delta^\sharp\mi\gamma
        +\mi m\cdot\delta^\sharp\mi\kappa
      \right)\d s .
  \label{eq:dictionary_internal_virtual_work}
\end{equation}
Its space--time counterpart is obtained by integration over \(\mathcal T\).

\subsection{Weak form, local balances, and initial-boundary conditions}
\label{subsec:dictionary_balance}

Let \(\mi f\) and \(\mi c\) denote prescribed distributed force and couple resultants per unit reference length, and let \(\bar{\mi t}\) and \(\bar{\mi m}\) denote prescribed end force and end couple resultants on \(\Gamma_r^N\) and \(\Gamma_R^N\), respectively.
Since \(\partial\mathfrak I=\{0,L\}\) is a zero-dimensional boundary, boundary integrals below are understood as sums over the corresponding beam ends.

Integrating the classical spatial weak form over \(\mathcal T\) gives the following space--time integral statement: find \((\mi r,\ten R)\in\mathcal Q_r\times\mathcal Q_R\) where
\[
  \mi p,\mi\ell\in H^1\bigl(\mathcal T;L^2(\mathfrak I;\R^3)\bigr),
  \qquad
  \mi n,\mi m\in L^2\bigl(\mathcal T;H^1(\mathfrak I;\R^3)\bigr),
\]
and such that, for all \((\mi\eta,\mi\theta)\in\mathcal V_r\times\mathcal V_\theta\),
\begin{align}
  &\int_{\mathcal B_0}
  \Big[
    \mi p_{,t}\cdot\mi\eta
    +\mi\ell_{,t}\cdot\mi\theta
    +\mi n\cdot\left(\mi\eta_{,s}-\mi\theta\times\mi r_{,s}\right)
    +\mi m\cdot\mi\theta_{,s}
    -\mi f\cdot\mi\eta
    -\mi c\cdot\mi\theta
  \Big]\d s\,\d t
  \nonumber\\
  &\hspace{3em}
  -\int_{\mathcal T}
  \left[
    \sum_{s_e\in\Gamma_r^N}\bar{\mi t}(s_e,t)\cdot\mi\eta(s_e,t)
    +\sum_{s_e\in\Gamma_R^N}\bar{\mi m}(s_e,t)\cdot\mi\theta(s_e,t)
  \right]\d t
  =0.
  \label{eq:dictionary_weak_balance}
\end{align}
The time derivatives of the momenta have not been integrated by parts in \eqref{eq:dictionary_weak_balance}; the equation is nevertheless tested and integrated over both \(s\) and \(t\).
Spatial integration by parts and localization with respect to the virtual fields give the local balance equations
\begin{equation}
  \mi p_{,t}=\mi n_{,s}+\mi f,
  \qquad
  \mi\ell_{,t}=\mi m_{,s}+\mi r_{,s}\times\mi n+\mi c
  \qquad\text{in }\mathcal B_0,
  \label{eq:dictionary_balance_equations}
\end{equation}
together with the spatial boundary conditions
\begin{align}
  \mi r&=\bar{\mi r}
  &&\text{on }\Sigma_r^D,
  &
  \nu_s\mi n&=\bar{\mi t}
  &&\text{on }\Sigma_r^N,
  \label{eq:dictionary_translational_boundary_conditions}
  \\
  \ten R&=\bar{\ten R}
  &&\text{on }\Sigma_R^D,
  &
  \nu_s\mi m&=\bar{\mi m}
  &&\text{on }\Sigma_R^N.
  \label{eq:dictionary_rotational_boundary_conditions}
\end{align}
The initial-value problem is completed by
\begin{equation}
  \mi r(s,0)=\mi r^{\mathrm{in}}(s),
  \qquad
  \ten R(s,0)=\ten R^{\mathrm{in}}(s),
  \qquad
  \mi p(s,0)=\mi p^{\mathrm{in}}(s),
  \qquad
  \mi\ell(s,0)=\mi\ell^{\mathrm{in}}(s).
  \label{eq:dictionary_initial_conditions}
\end{equation}
Equivalently, the last two conditions prescribe the initial translational and angular velocities through \eqref{eq:dictionary_temporal_momenta}. No terminal condition is imposed at \(t=t_{\mathrm{end}}\).
% Version 8 (26 October 2026): integrate the additional row-reduction details from Appendix A.1.
\section{The beam as a directed world sheet in space--time}
\label{sec:world_sheet}

The purpose of this section is to formulate the balance equations of Section~\ref{sec:cosserat_dictionary}, including its curved, stress-free reference configurations, directly on a parametrized two-dimensional world sheet in non-relativistic space--time.
The construction is shell-like in the sense that the beam history is treated as an immersed surface equipped with directors and co-normal boundary fluxes.
It is, however, not a relativistic model: time remains absolute and only spatial rotations are admissible.

\subsection{Non-relativistic space--time parametrization}
\label{subsec:worldsheet_embedding}

We use the product space
\[
  \R^4\cong\R^3\times\R
\]
with canonical basis \(\{\mi e_i\}_{i=1}^4\).
The first three basis vectors span physical space, while \(\mi e_4\) represents the temporal direction.
The centerline history is parametrized by
\begin{equation}
  \Xi:\mathcal B_0\rightarrow\R^4,
  \qquad
  (s,t)\mapsto
  \Xi(s,t):=
  \begin{bmatrix}
    \mi r(s,t)\\[0.2em]
    \tau(s,t)
  \end{bmatrix}.
  \label{eq:worldsheet_embedding}
\end{equation}
In classical mechanics time is not an additional deformation variable.
We therefore restrict admissible configurations by
\begin{equation}
  \tau(s,t)=c_\tau t,
  \qquad c_\tau>0,
  \label{eq:worldsheet_time_constraint}
\end{equation}
Here \(c_\tau\) is a fixed conversion factor from time to length units, not a material wave speed.
Consequently,
\begin{equation}
  \delta\tau=0,
  \qquad
  \delta\Xi=\begin{bmatrix}\mi\eta\\0\end{bmatrix}.
  \label{eq:worldsheet_admissible_variation}
\end{equation}
The temporal coordinate is therefore considered as prescribed geometric data. The image
\[
  \mathcal W:=\Xi(\mathcal B_0)
\]
is called the directed world sheet: the centerline history represented as a parametrized surface.
The coordinates \((s,t)\), with \(s\) the reference arc length, remain its independent parameters even for a curved reference centerline.
Its tangent vectors are
\begin{equation}
  \mi g_s:=\Xi_{,s}=\begin{bmatrix}\mi r_{,s}\\0\end{bmatrix},
  \qquad
  \mi g_t:=\Xi_{,t}=\begin{bmatrix}\mi r_{,t}\\c_\tau\end{bmatrix}.
  \label{eq:worldsheet_tangents}
\end{equation}
The induced metric and surface measure are
\begin{equation}
  g_{\mu\nu}:=\mi g_\mu\cdot\mi g_\nu,
  \qquad
  \d A_{st}=\sqrt g\,\d s\,\d t,
  \qquad
  g:=\det(g_{\mu\nu}),
  \label{eq:worldsheet_metric}
\end{equation}
where \(\mu,\nu\in\{s,t\}\); repeated surface indices are summed over these two directions.
Explicitly,
\begin{equation}
  g_{ss}=\mi r_{,s}\cdot\mi r_{,s},
  \qquad
  g_{st}=\mi r_{,s}\cdot\mi r_{,t},
  \qquad
  g_{tt}=\mi r_{,t}\cdot\mi r_{,t}+c_\tau^2.
  \label{eq:worldsheet_metric_components}
\end{equation}
Using \(\det(g_{\mu\nu})=g_{ss}g_{tt}-g_{st}^{2}\), we obtain
\[
  g=c_\tau^2\|\mi r_{,s}\|^2
    +\|\mi r_{,s}\times\mi r_{,t}\|^2.
\]
For the geometric construction we assume sufficiently smooth histories, with \(\mi r\) at least \(C^2\) on each regular patch and
\begin{equation}
  \|\mi r_{,s}\|\ge c_r>0.
  \label{eq:worldsheet_nondegeneracy}
\end{equation}
Thus \(g\ge c_\tau^2c_r^2>0\), and \(\Xi\) is a regular immersion.
These regularity and nondegeneracy assumptions are additional to the \(H^1\) configuration-space definitions.
Global injectivity is not assumed: if different material points occupy the same spatial position at the same time, their material labels are retained.
Surface and boundary integrals are understood through the pull-back to the corresponding parameter-domain patches, with material multiplicity at self-intersections.

The induced metric is used below only to define intrinsic divergence, surface measures, and co-normal boundary fluxes.
The material beam energetics remain those of Section~\ref{sec:cosserat_dictionary} and enter the world-sheet formulation through the variational principle in Subsection~\ref{subsec:worldsheet_weak_fluxes}.
The conversion factor \(c_\tau\) affects this auxiliary geometry; it cancels from the material balance equations and the boundary virtual actions after pull-back.

\subsection{Embedded director frame}
\label{subsec:worldsheet_directors}

For a spatial vector \(\mi a\in\R^3\) we define the canonical lift
\[
  \hat{\mi a}:=\begin{bmatrix}\mi a\\0\end{bmatrix}\in\R^4.
\]
The lift appends a zero temporal component; no quaternion multiplication or parametrization is used.
The beam directors are lifted by
\begin{equation}
  \hat{\mi d}_i:=\begin{bmatrix}\mi d_i\\0\end{bmatrix},
  \qquad
  \hat{\mi D}_i:=\begin{bmatrix}\mi D_i\\0\end{bmatrix},
  \qquad
  i=1,2,3,
  \qquad
  \hat{\mi d}_4=\hat{\mi D}_4:=\mi e_4.
  \label{eq:worldsheet_lifted_directors}
\end{equation}
The reference directors \(\mi D_i(s)\) may vary along the beam but are fixed in time and under mechanical variations.
The associated embedded relative rotation is
\begin{equation}
  \hat{\ten R}
  :=\hat{\mi d}_a\otimes\hat{\mi D}_a
  =
  \begin{bmatrix}
    \ten R&0\\
    0&1
  \end{bmatrix}
  \in\SO{}(4),
  \qquad a=1,2,3,4.
  \label{eq:worldsheet_hatR}
\end{equation}
The sum over all four directors gives the canonical embedding of \(\SO{}(3)\) into \(\SO{}(4)\).
The constraint \(\hat{\ten R}\mi e_4=\mi e_4\) excludes rotations that mix spatial and temporal directions.
Admissible rotational variations have the same block structure,
\begin{equation}
  \delta\hat{\ten R}=\hat{\ten\Theta}\hat{\ten R},
  \qquad
  \hat{\ten\Theta}:=
  \begin{bmatrix}
    \VPTen{\mi\theta}&0\\
    0&0
  \end{bmatrix}.
  \label{eq:worldsheet_hatR_variation}
\end{equation}
Hence admissible rotational variations still have only three independent components, represented by the spatial virtual spin \(\mi\theta\).
The spatial angular velocity and the spatial elastic curvature change remain those of Section~\ref{sec:cosserat_dictionary}, namely
\[
  \mi\omega=\axl(\ten R_{,t}\ten R\tp),
  \qquad
  \mi\kappa=\axl(\ten R_{,s}\ten R\tp).
\]
Here \(\mi\kappa=\ten R\mathcal K\) is not the absolute curvature of the current triad: differentiation of \(\mi d_i=\ten R\mi D_i\) gives
\[
  \mi d_{i,s}=\mi\kappa\times\mi d_i+\ten R\mi D_{i,s},
  \qquad
  \mi d_{i,t}=\mi\omega\times\mi d_i.
\]
The reference-curvature contribution is already accounted for by the relative rotation, as shown in Subsection~\ref{subsec:dictionary_strains}.

\subsection{Space--time fluxes}
\label{subsec:worldsheet_fluxes}

Let \(\xi^1=s\), \(\xi^2=t\).
For an \(\R^3\)-valued surface flux \(\mathfrak q^\mu\), contravariant in the surface index \(\mu\in\{s,t\}\), we use the covariant divergence
\begin{equation}
  \mathfrak q^\mu{}_{;\mu}
  =\frac1{\sqrt g}\left(\sqrt g\,\mathfrak q^\mu\right)_{,\mu}.
  \label{eq:worldsheet_divergence}
\end{equation}
This divergence acts on the surface index, componentwise in a fixed spatial Cartesian basis. The test fields \(\mi\eta\) and \(\mi\theta\) carry no surface index; hence \(\mi\eta_{;\mu}=\mi\eta_{,\mu}\) and \(\mi\theta_{;\mu}=\mi\theta_{,\mu}\).
In the distinguished material coordinates \((s,t)\), the spatial resultants \(\mi n,\mi m\) and the temporal momenta \(\mi p,\mi\ell\) are combined into the space--time fluxes
\begin{equation}
  \sqrt g\,\mathfrak n^s:=\mi n,
  \qquad
  \sqrt g\,\mathfrak n^t:=-\mi p,
  \qquad
  \sqrt g\,\mathfrak m^s:=\mi m,
  \qquad
  \sqrt g\,\mathfrak m^t:=-\mi\ell.
  \label{eq:worldsheet_flux_definition}
\end{equation}
The minus sign in the temporal components is the sign convention that turns the local momentum balances into divergence equations.
Under a change of surface coordinates, \(\mathfrak n^\mu\) and \(\mathfrak m^\mu\) transform contravariantly in the surface index; the products with \(\sqrt g\) transform as vector densities.
The corresponding source terms are
\begin{equation}
  \tilde{\mi f}:=\frac{1}{\sqrt g}\mi f,
  \qquad
  \tilde{\mi s}:=\frac{1}{\sqrt g}\left(\mi r_{,s}\times\mi n+\mi c\right).
  \label{eq:worldsheet_source_definition}
\end{equation}
Then the first-order balance equations \eqref{eq:dictionary_balance_equations} are equivalently written as
\begin{equation}
  \mathfrak n^\mu{}_{;\mu}+\tilde{\mi f}=\mi0,
  \qquad
  \mathfrak m^\mu{}_{;\mu}+\tilde{\mi s}=\mi0
  \qquad\text{on }\mathcal W.
  \label{eq:worldsheet_covariant_balance}
\end{equation}
Indeed, using \eqref{eq:worldsheet_divergence} and \eqref{eq:worldsheet_flux_definition}, the first equation gives
\[
  \frac1{\sqrt g}\left(\mi n_{,s}-\mi p_{,t}+\mi f\right)=\mi0,
\]
and the second gives
\[
  \frac1{\sqrt g}\left(\mi m_{,s}-\mi\ell_{,t}+\mi r_{,s}\times\mi n+\mi c\right)=\mi0.
\]
Thus the original momentum balances are expressed in terms of force and moment fluxes on the world sheet.
The reference orientation and section data enter through the resultants and their spatial derivatives, not through an additional connection on the spatial vector components.
The co-normal is dual to the outward unit vector tangent to the surface and normal to its boundary curve.

\subsection{Variational principle, weak form, and co-normal boundary fluxes}
\label{subsec:worldsheet_weak_fluxes}

The geometric construction is now coupled to the beam energetics introduced in Section~\ref{sec:cosserat_dictionary}.
With the kinetic-energy density \(K\) from \eqref{eq:dictionary_kinetic_energy} and the stored-energy density \(\Psi\) from \eqref{eq:dictionary_energy}, the Lagrangian density per unit material length is
\begin{equation}
  \mathcal L
  :=K-\Psi.
  \label{eq:worldsheet_lagrangian_density}
\end{equation}
The action of the beam is
\begin{equation}
  \mathcal A[\mi r,\ten R]
  :=\int_{\mathcal B_0}\mathcal L\,\d s\,\d t.
  \label{eq:worldsheet_action_parameter_domain}
\end{equation}
All physical densities used here are defined with respect to the material space--time measure \(\d s\,\d t\).
Their representation on the embedded world sheet is therefore only a change of measure:
\begin{equation}
  \widetilde{\mathcal L}:=\frac{\mathcal L}{\sqrt g},
  \qquad
  \mathcal A[\mi r,\ten R]
  =\int_{\mathcal W}\widetilde{\mathcal L}\,\d A_{st}.
  \label{eq:worldsheet_change_of_measure}
\end{equation}
No surface-tension energy is introduced by \eqref{eq:worldsheet_change_of_measure}; the metric factor merely converts material densities into surface densities on \(\mathcal W\).
Variations of the surface representation are understood through the pull-back to \(\mathcal B_0\), so that the variation of \(\sqrt g\) in the area measure is exactly compensated by its occurrence in \(\widetilde{\mathcal L}\).

For distributed forces and couples, we postulate the balance of virtual actions in Lagrange--d'Alembert form,
\begin{equation}
  \delta\mathcal A+\delta\mathcal A_{\mathrm{ext}}=0
  \qquad
  \forall\,(\mi\eta,\mi\theta)\in\mathcal V_r\times\mathcal V_\theta,
  \label{eq:worldsheet_lagrange_dalembert}
\end{equation}
where
\begin{equation}
  \delta\mathcal A_{\mathrm{ext}}
  :=\int_{\mathcal B_0}
      \left(
        \mi f\cdot\mi\eta
        +\mi c\cdot\mi\theta
      \right)\d s\,\d t
    +\delta\mathcal A^{\mathrm{nat}}_{\partial\mathcal W}.
  \label{eq:worldsheet_external_virtual_work}
\end{equation}
Here \(\delta\mathcal A_{\mathrm{ext}}\) denotes an external virtual-action functional, linear in \((\mi\eta,\mi\theta)\); the notation does not require the existence of an external potential.
Its boundary part \(\delta\mathcal A^{\mathrm{nat}}_{\partial\mathcal W}\) initially denotes the as-yet unspecified virtual action of boundary interactions.
As in spatial virtual-work formulations, the conjugate boundary quantities are identified by integrating \(\delta\mathcal A\) by parts.
Applying this construction in both \(s\) and \(t\) determines the structure of the spatial and temporal boundary contributions below, but not the values of the applied loads.
The test spaces were defined in \eqref{eq:dictionary_test_space_r} and \eqref{eq:dictionary_test_space_theta}; in particular, the virtual fields vanish on the spatial Dirichlet boundaries and on the initial face, but not on the terminal face.

Using the kinetic variation \eqref{eq:dictionary_kinetic_variation} and the internal virtual work \eqref{eq:dictionary_internal_virtual_work}, the first variation of the action is
\begin{align}
  \delta\mathcal A
  =\int_{\mathcal B_0}
  \Big[
    \mi p\cdot\mi\eta_{,t}
    +\mi\ell\cdot\mi\theta_{,t}
    -\mi n\cdot\delta^\sharp\mi\gamma
    -\mi m\cdot\delta^\sharp\mi\kappa
  \Big]\d s\,\d t.
  \label{eq:worldsheet_action_variation}
\end{align}
Substitution of the pushed-forward material variations \eqref{eq:dictionary_pushed_variations}, followed by the scalar triple-product identity, gives
\begin{align}
  0=
  &\int_{\mathcal B_0}
  \Big[
    \mi p\cdot\mi\eta_{,t}
    +\mi\ell\cdot\mi\theta_{,t}
    -\mi n\cdot\mi\eta_{,s}
    -\mi m\cdot\mi\theta_{,s}
    +\mi f\cdot\mi\eta
    +\left(\mi r_{,s}\times\mi n+\mi c\right)\cdot\mi\theta
  \Big]\d s\,\d t
  +\delta\mathcal A^{\mathrm{nat}}_{\partial\mathcal W}.
  \label{eq:worldsheet_weak_parameter_form}
\end{align}
Using the change of measure \eqref{eq:worldsheet_change_of_measure} and the flux definitions \eqref{eq:worldsheet_flux_definition}--\eqref{eq:worldsheet_source_definition}, this statement localizes on each regular material patch \(\mathcal B_e\subset\mathcal B_0\), with parametrized image \(\mathcal W_e:=\Xi(\mathcal B_e)\), as
\begin{align}
  0=
  \mathcal R_{\mathcal W_e}(\mi\eta,\mi\theta)
  :=
  \int_{\mathcal W_e}
  \Big[
    -\mathfrak n^\mu\cdot\mi\eta_{;\mu}
    -\mathfrak m^\mu\cdot\mi\theta_{;\mu}
    +\tilde{\mi f}\cdot\mi\eta
    +\tilde{\mi s}\cdot\mi\theta
  \Big] \d A_{st}
  +\delta\mathcal A^{\mathrm{nat}}_{\partial\mathcal W_e}.
  \label{eq:worldsheet_weak_flux_form}
\end{align}
On artificial patch boundaries, the boundary interactions are the co-normal traces of the adjacent material patches, with opposite outward orientations.

Let \(\lambda=\lambda_\mu\d\xi^\mu\) denote the outward unit co-normal covector on \(\partial\mathcal W_e\), and let \(\d\ell\) be the induced line measure.
The surface divergence theorem yields
\begin{align}
  \mathcal R_{\mathcal W_e}
  =&\int_{\mathcal W_e}
  \Big[
    \left(\mathfrak n^\mu{}_{;\mu}+\tilde{\mi f}\right)\cdot\mi\eta
    +\left(\mathfrak m^\mu{}_{;\mu}+\tilde{\mi s}\right)\cdot\mi\theta
  \Big]\d A_{st}
  \nonumber\\
  &-
  \int_{\partial\mathcal W_e}
  \Big[
    \left(\mathfrak n^\mu\lambda_\mu\right)\cdot\mi\eta
    +\left(\mathfrak m^\mu\lambda_\mu\right)\cdot\mi\theta
  \Big]\d\ell
  +\delta\mathcal A^{\mathrm{nat}}_{\partial\mathcal W_e}.
  \label{eq:worldsheet_divergence_theorem_weak}
\end{align}
Localization in the interior and on boundary parts with free test traces separates the bulk balances from the boundary work.
The coefficients of \(\mi\eta\) and \(\mi\theta\) in the boundary integral identify the conjugate co-normal fluxes
\begin{equation}
  \mathfrak t_r:=\mathfrak n^\mu\lambda_\mu,
  \qquad
  \mathfrak t_\theta:=\mathfrak m^\mu\lambda_\mu.
  \label{eq:worldsheet_conormal_flux}
\end{equation}
This single expression covers spatial beam ends, temporal faces of a slab, and non-aligned edges of a locally refined space--time mesh.
To distinguish the geometric and material boundary measures, orient the material chart by \(\d s\wedge\d t\) and denote its outward boundary-density components by \(\nu_\mu\,\d\widehat\ell\).
Along the positively oriented boundary of \(\mathcal B_e\), they satisfy
\[
  \nu_s\,\d\widehat\ell=\d t,
  \qquad
  \nu_t\,\d\widehat\ell=-\d s.
\]
Only these density--measure products are needed; \(\d\widehat\ell\) is the line measure in the fixed parameter chart, whereas \(\d\ell\) is induced by the world-sheet metric.
The boundary pull-back obeys
\begin{equation}
  \Xi^*\!\left(\lambda_\mu\,\d\ell\right)
  =\sqrt g\,\nu_\mu\,\d\widehat\ell.
  \label{eq:worldsheet_conormal_pullback}
\end{equation}
Consequently, the flux--measure products take the material form
\begin{align}
  \Xi^*\!\left[
    \left(\mathfrak n^\mu\lambda_\mu\right)\d\ell
  \right]
  &=\left(\mi n\nu_s-\mi p\nu_t\right)\d\widehat\ell,
  \nonumber\\
  \Xi^*\!\left[
    \left(\mathfrak m^\mu\lambda_\mu\right)\d\ell
  \right]
  &=\left(\mi m\nu_s-\mi\ell\nu_t\right)\d\widehat\ell.
  \label{eq:worldsheet_material_boundary_fluxes}
\end{align}
Thus material flux densities are paired with \(\d\widehat\ell\), and geometric co-normal flux densities with \(\d\ell\); both representations give the same boundary virtual action.

Let \(\partial\mathcal W_{N,r}\) and \(\partial\mathcal W_{N,\theta}\) denote the natural boundary parts for translation and rotation, respectively.
Their lateral parts are the images of \(\Sigma_r^N\) and \(\Sigma_R^N\); temporal boundary interactions are included wherever the corresponding test traces are free.
The natural boundary virtual action is
\begin{equation}
  \delta\mathcal A^{\mathrm{nat}}_{\partial\mathcal W}
  =\int_{\partial\mathcal W_{N,r}}
      \bar{\mathfrak t}_r\cdot\mi\eta\,\d\ell
   +\int_{\partial\mathcal W_{N,\theta}}
      \bar{\mathfrak t}_\theta\cdot\mi\theta\,\d\ell,
  \label{eq:worldsheet_natural_virtual_work}
\end{equation}
so that the weak statement enforces \(\mathfrak t_r=\bar{\mathfrak t}_r\) on \(\partial\mathcal W_{N,r}\) and \(\mathfrak t_\theta=\bar{\mathfrak t}_\theta\) on \(\partial\mathcal W_{N,\theta}\).
For a coordinate-aligned slab \(\mathcal B_n=\mathfrak I\times(t_n,t_{n+1})\), the co-normal fluxes reduce to the familiar physical quantities.
With the outward orientation of the slab one obtains
\begin{center}
\begin{tabular}{c|c|c}
  boundary face & \((\mathfrak n^\mu\lambda_\mu)\d\ell\) & \((\mathfrak m^\mu\lambda_\mu)\d\ell\) \\
  \hline
  \(s=L\) & \(\phantom{-}\mi n\,\d t\) & \(\phantom{-}\mi m\,\d t\) \\
  \(s=0\) & \(-\mi n\,\d t\) & \(-\mi m\,\d t\) \\
  \(t=t_{n+1}\) & \(-\mi p\,\d s\) & \(-\mi\ell\,\d s\) \\
  \(t=t_n\) & \(\phantom{-}\mi p\,\d s\) & \(\phantom{-}\mi\ell\,\d s\)
\end{tabular}
\end{center}
Consequently, the natural virtual work on the temporal faces has the form
\begin{align}
  \delta\mathcal A^{\mathrm{nat}}_{t_n,t_{n+1}}
  =&-
  \int_{\mathfrak I}
    \left(
      \bar{\mi p}^{+}\cdot\mi\eta
      +\bar{\mi\ell}^{+}\cdot\mi\theta
    \right)(s,t_{n+1})\d s
  \nonumber\\
  &+
  \int_{\mathfrak I}
    \left(
      \bar{\mi p}^{-}\cdot\mi\eta
      +\bar{\mi\ell}^{-}\cdot\mi\theta
    \right)(s,t_n)\d s .
  \label{eq:worldsheet_temporal_natural_work}
\end{align}
These are the temporal analogues of spatial Neumann tractions: they represent momentum transfer through the temporal cuts, not an additional applied force.
For the initial-value problem, the initial data are prescribed, whereas \(\bar{\mi p}^{+}\) and \(\bar{\mi\ell}^{+}\) are identified with the interior outgoing traces of the solution, not with independently prescribed terminal values.
At the physical initial face, \(\bar{\mi p}^{-}=\mi p^{\mathrm{in}}\) and \(\bar{\mi\ell}^{-}=\mi\ell^{\mathrm{in}}\).
Since the unperturbed configuration tests vanish there, the initial momentum conditions in \eqref{eq:dictionary_initial_conditions} are imposed separately as admissibility conditions; they do not follow from localization of the vanishing lower boundary work.
The final time \(t_{\mathrm{end}}\) is fixed throughout, while the terminal configuration is free.
In \eqref{eq:worldsheet_temporal_natural_work}, these momenta enter as coefficients of a boundary virtual-action functional evaluated on \((\mi\eta,\mi\theta)\), rather than through an additional boundary potential.
If the upper temporal flux term is omitted while the final configuration is not prescribed, the variational problem imposes the artificial terminal condition \(\mi p=\mi0\), \(\mi\ell=\mi0\) on that face.
The outgoing-trace identification makes the boundary-completed weak statement equivalent to the classical initial-value balances without adding terminal data.

\subsection{Mixed first-order form for the subsequent discretization}
\label{subsec:worldsheet_mixed_form}

The momenta \(\mi p\) and \(\mi\ell\) are introduced as independent fields directly in the continuous space--time action, before discretization.
Following the classical Legendre construction of analytical dynamics~\cite[Sec.~26.2]{pars_treatise_1965}, we transform only the temporal rates; the spatial strain measures remain in \(\Psi\).
Starting from the Lagrangian density \eqref{eq:worldsheet_lagrangian_density}, we define
\begin{equation}
  \mathcal H
  :=\mi p\cdot\mi r_{,t}
    +\mi\ell\cdot\mi\omega
    -\mathcal L.
  \label{eq:worldsheet_legendre_transform}
\end{equation}
Elimination of the temporal rates gives
\begin{equation}
  \mathcal H
  =\frac12 A^{-1}\mi p\cdot\mi p
   +\frac12 \mi\ell\cdot(\ten J^{\mathrm{sp}})^{-1}\mi\ell
   +\Psi(s,\mi\Gamma,\mathcal K).
  \label{eq:worldsheet_hamiltonian_density}
\end{equation}
The density \(\mathcal H\) depends on \(s,\ten R,\mi r_{,s},\ten R_{,s},\mi p,\mi\ell\); its spatial-derivative dependence is carried by \(\Psi(s,\mi\Gamma,\mathcal K)\).
The explicit \(s\)-dependence includes the reference and section data from Section~\ref{sec:cosserat_dictionary}; in particular, \(\ten J^{\mathrm{sp}}=\ten R\ten J(s)\ten R\tp\).
The associated phase-space action is
\begin{equation}
  \mathcal A_{\mathrm{mix}}[\mi r,\ten R,\mi p,\mi\ell]
  :=\int_{\mathcal B_0}
    \left(
      \mi p\cdot\mi r_{,t}
      +\mi\ell\cdot\mi\omega
      -\mathcal H
    \right)\d s\,\d t.
  \label{eq:worldsheet_mixed_action}
\end{equation}
At the level of the action density, this construction is related to Hamilton--Pontryagin formulations~\cite{yoshimura_reduction_2007}: momenta enforce compatibility between independently varied velocities and configuration rates.
For the present quadratic kinetic energy, the momentum--velocity relations are invertible; eliminating the auxiliary velocities yields the density in \eqref{eq:worldsheet_mixed_action}.
For the corresponding relation to Livens' principle in space--time rigid-body dynamics, see~\cite[Appendix~A]{hesch_space-time_2024}.
Here the variational problem retains the external virtual action and temporal co-normal terms of Subsection~\ref{subsec:worldsheet_weak_fluxes}; a fixed terminal configuration is not assumed.

Stationarity of \(\mathcal A_{\mathrm{mix}}\) with respect to \(\mi p\) and \(\mi\ell\) gives the temporal constitutive relations
\begin{equation}
  \mi r_{,t}=A^{-1}\mi p,
  \qquad
  \mi\omega=(\ten J^{\mathrm{sp}})^{-1}\mi\ell.
  \label{eq:worldsheet_temporal_constitutive_relations}
\end{equation}
The configuration fields belong to \(\mathcal Q_r\times\mathcal Q_R\).
For the regular continuous formulation considered here, the independent momenta retain the temporal regularity
\[
  \mi p,\mi\ell\in H^1\bigl(\mathcal T;L^2(\mathfrak I;\R^3)\bigr)
\]
and satisfy the prescribed initial traces in \eqref{eq:dictionary_initial_conditions}.
The momentum variations \(\mi\pi=\delta\mi p\) and \(\mi\rho=\delta\mi\ell\) belong to the corresponding homogeneous spaces: their initial traces vanish and their terminal traces are unrestricted.

The discretization in Subsection~\ref{subsec:disc_galerkin_form}
starts from a continuous mixed system in which the angular
momentum--velocity relation
\eqref{eq:worldsheet_temporal_constitutive_relations}
is used to simplify the rotational equation.
The following calculation shows why this preserves the regular
continuous solutions when the constitutive relation is retained
as an independent equation.

Define the residual of this constitutive relation by
\[
  \mathcal E_\ell
  :=\mi\omega-(\ten J^{\mathrm{sp}})^{-1}\mi\ell.
\]
In the variation with respect to \(\ten R\), \(\mi\ell\) and the
reference data are held fixed while both \(\mi\omega\) and
\(\ten J^{\mathrm{sp}}\) vary, giving
\begin{equation}
  \delta_{\ten R}\left(
    \mi\ell\cdot\mi\omega
    -\frac12\mi\ell\cdot(\ten J^{\mathrm{sp}})^{-1}\mi\ell
  \right)
  =\mi\ell\cdot\mi\theta_{,t}
   +\mi\theta\cdot(\mathcal E_\ell\times\mi\ell).
  \label{eq:worldsheet_rotational_phase_variation}
\end{equation}
The additional term need not vanish for arbitrary independent
fields, but it vanishes when the constitutive relation is
satisfied. Its dependence on the constitutive residual follows
from
\(
  \mi\theta\cdot(\mathcal E_\ell\times\mi\ell)
  =\mathcal E_\ell\cdot(\mi\ell\times\mi\theta)
\).

For \(\mathbf y:=(\mi r,\ten R,\mi p,\mi\ell)\), let
\(\mathcal R_\theta^{\mathrm{act}}\) denote the rotational
residual obtained by direct variation of the mixed action
together with the external virtual action.
Let \(\mathcal R_\theta^{\mathrm{red}}\) denote the same residual
with the additional term omitted, as used in
\eqref{eq:worldsheet_mixed_weak_form}.
Then
\begin{equation}
  \mathcal R_\theta^{\mathrm{act}}(\mathbf y;\mi\theta)
  =\mathcal R_\theta^{\mathrm{red}}(\mathbf y;\mi\theta)
   +\int_{\mathcal B_0}
    \mi\theta\cdot(\mathcal E_\ell\times\mi\ell)\,\d s\,\d t.
  \label{eq:worldsheet_action_row_relation}
\end{equation}
For a sufficiently regular continuous solution, the independent
momentum variation enforces \(\mathcal E_\ell=\mi0\) in the
interior.
The homogeneous initial traces of the momentum variations do
not obstruct this conclusion, since compactly supported
variations remain admissible.
Retaining the constitutive equation therefore allows us to
replace \(\mathcal R_\theta^{\mathrm{act}}=0\) by
\(\mathcal R_\theta^{\mathrm{red}}=0\) without changing the
continuous solution set.

This simplification is performed before discretization;
it does not imply that discretizing the unreduced mixed action
yields an equivalent discrete system.
The consequences for the discrete test spaces are discussed in
Subsection~\ref{subsec:disc_galerkin_form}.

The form used for the subsequent discretization is therefore: find \((\mi r,\ten R,\mi p,\mi\ell)\) such that for all admissible \((\mi\eta,\mi\theta,\mi\pi,\mi\rho)\)
\begin{align}
  0=&
  \int_{\mathcal B_0}
  \Big[
    \mi p\cdot\mi\eta_{,t}
    +\mi\ell\cdot\mi\theta_{,t}
    -\mi n\cdot\delta^\sharp\mi\gamma
    -\mi m\cdot\delta^\sharp\mi\kappa
    +\mi f\cdot\mi\eta
    +\mi c\cdot\mi\theta
  \Big]\d s\,\d t
  \nonumber\\
  &+
  \int_{\mathcal B_0}
  \Big[
    (\mi r_{,t}-A^{-1}\mi p)\cdot\mi\pi
    +(\mi\omega-(\ten J^{\mathrm{sp}})^{-1}\mi\ell)\cdot\mi\rho
  \Big]\d s\,\d t
  +\delta\mathcal A^{\mathrm{nat}}_{\partial\mathcal W}.
  \label{eq:worldsheet_mixed_weak_form}
\end{align}
Here \(\delta^\sharp\mi\gamma\) and \(\delta^\sharp\mi\kappa\) are the pushed-forward material variations from \eqref{eq:dictionary_pushed_variations}. After integration by parts it recovers the flux balances \eqref{eq:worldsheet_covariant_balance} together with the temporal constitutive equations \eqref{eq:worldsheet_temporal_constitutive_relations}.

\begin{remark}
The angular momentum \(\mi\ell\) in \eqref{eq:worldsheet_flux_definition} and \eqref{eq:worldsheet_mixed_weak_form} is spatial.
With this convention the total angular momentum density is \(\mi r\times\mi p+\mi\ell\).
For comparison, a referential momentum \(\ten R\tp\mi\ell\) is pushed forward by \(\ten R\), whereas local director components \(\ten\Lambda\tp\mi\ell\) are pushed forward by \(\ten\Lambda=\ten R\ten\Lambda_0\). The formulation above retains spatial momenta.
\end{remark}

\subsection{Time orientation and preparation for stabilization}
\label{subsec:worldsheet_time_orientation}

The world sheet is not merely an unoriented surface in \(\R^4\).
It carries a distinguished future direction inherited from the absolute time coordinate.
In the aligned coordinates used above this direction is represented by \(\partial_t\).
On a general space--time chart with coordinates \(\xi^\mu\), the absolute time is a scalar field \(t=t(\xi)\).
A vector field is future-directed when
\[
  \beta=\beta^\mu\partial_\mu,
  \qquad
  \d t(\beta)=\beta^\mu t_{,\mu}>0.
\]
This reduces to \(\beta^t>0\) in a chart containing \(t\) as a coordinate.
For a field \(z\) on the world sheet we write
\[
  D_\beta z:=\beta^\mu z_{;\mu}.
\]
The stabilized discretization introduced later uses this operator in the Petrov--Galerkin test modification
\[
  z_h^\star=z_h+\tau_e D_\beta z_h,
\]
with an elementwise stabilization time scale \(\tau_e\), distinct from the prescribed embedding coordinate \(\tau=c_\tau t\).
For the aligned choice \(\beta=\partial_t\), \(D_\beta\) is the partial time derivative at fixed \(s\).
In any chart, this same material time direction is characterized by \(\beta(s)=0\) and \(\beta(t)=1\); a general future-directed field need not have this normalization or preserve the material label \(s\).
For non-aligned space--time meshes, the same notation identifies the facet completion required by the future-directed test perturbation; the recommended numerical method below retains the absolute-time-aligned tensor-product structure.

% Version 8 (26 October 2026): point the Galerkin construction to the integrated continuous row reduction.
%============================================================
\section{Future-directed stabilized tensor-product space--time discretization}
\label{sec:spacetime_discretization}

The world-sheet balance laws of Section~\ref{sec:world_sheet} do not prescribe a particular space--time mesh.
For the numerical method developed and recommended in this paper, however, we retain the distinguished foliation of non-relativistic mechanics and use quadrilateral elements aligned with the material coordinate and absolute time.
This choice retains the prescribed time foliation and leads to the one-parameter future-directed Petrov--Galerkin method below, without an additional interior numerical-flux parameter.
A conservative extension to non-aligned facets is recorded in Appendix~\ref{subsec:app_non_aligned_facets}; the numerical examples show that formal consistency of that extension does not guarantee mesh-uniform stability of continuous equal-order simplex spaces.

The mixed equations are discretized in an integral space--time form: temporal derivatives remain on the momentum fields, whereas the spatial stress-resultant terms have been integrated by parts.
The temporal integration by parts is combined with the incoming and outgoing co-normal terms of Subsection~\ref{subsec:worldsheet_weak_fluxes}; it does not introduce a prescribed terminal state.
All four test fields are then perturbed in the future direction \(\partial_t\). Neither this rearrangement nor the test perturbation imposes the equations pointwise in time.

%------------------------------------------------------------
\subsection{Time-aligned mesh and approximation spaces}
\label{subsec:disc_mesh_spaces}

Let
\[
  0=s_0<s_1<\cdots<s_{N_s}=L,
  \qquad
  0=t_0<t_1<\cdots<t_{N_t}=t_{\mathrm{end}},
\]
and define
\[
  I_a:=(s_a,s_{a+1}),
  \qquad
  T_n:=(t_n,t_{n+1}),
  \qquad
  \mathcal B_{an}:=I_a\times T_n.
\]
The material space--time cylinder is partitioned as
\begin{equation}
  \mathcal T_h^{\mathrm{st}}
  :=\left\{\mathcal B_{an}\right\}_{a=0,n=0}^{N_s-1,N_t-1},
  \qquad
  \overline{\mathcal B_0}
  =\bigcup_{a,n}\overline{\mathcal B_{an}}.
  \label{eq:disc_mesh}
\end{equation}
We use continuous tensor-product bases of degrees \((k_s,k_t)\) for the placement, rotation increments and their tests, and \((m_s,m_t)\) for the independent momenta and their tests.
The local time-step size is \(\Delta t_n:=t_{n+1}-t_n\); the equal-order quadratic choice \((k_s,k_t)=(m_s,m_t)=(2,2)\) is denoted by \(Q_2/Q_2\).
Prescribed initial traces of all four fields and spatial essential data are imposed strongly, with homogeneous traces for the corresponding test fields; the terminal boundary remains free.

%------------------------------------------------------------
\subsection{Discrete fields and group-valued kinematics}
\label{subsec:disc_fields}

Let \(\mathcal I_q\) and \(\mathcal I_p\) index all global scalar basis functions \(N_A\) for placement and rotation increments, and \(M_B\) for momenta, respectively.
The subsets \(\mathcal I_r^{\mathrm f},\mathcal I_R^{\mathrm f}\subset\mathcal I_q\) and \(\mathcal I_p^{\mathrm f}\subset\mathcal I_p\) identify the free placement, rotation-increment and momentum coefficients.
The placement and its test field are
\begin{equation}
  \mi r^h=\sum_{A\in\mathcal I_q}N_A\mi q_A,
  \qquad
  \mi\eta^h=\sum_{A\in\mathcal I_r^{\mathrm f}}N_A\mi\eta_A.
  \label{eq:disc_r_eta}
\end{equation}
The temporal coordinate of the embedding remains exact,
\begin{equation}
  \Xi^h(s,t)
  =\begin{bmatrix}\mi r^h(s,t)\\c_\tau t\end{bmatrix},
  \qquad
  \delta\Xi^h
  =\begin{bmatrix}\mi\eta^h\\0\end{bmatrix}.
  \label{eq:disc_Xi}
\end{equation}
At Newton iteration \(j\), the rotational increment and test field are
\begin{equation}
  \Delta\mi\vartheta^h
  =\sum_{A\in\mathcal I_R^{\mathrm f}}N_A\Delta\mi\vartheta_A,
  \qquad
  \mi\theta^h
  =\sum_{A\in\mathcal I_R^{\mathrm f}}N_A\mi\theta_A,
  \label{eq:disc_rot_increment}
\end{equation}
with the multiplicative Lie-group update \cite{iserles_lie-group_2000}
\begin{equation}
  \ten R^{h,j+1}
  =\operatorname{exp}{\VPTen{\Delta\mi\vartheta^h}}\ten R^{h,j},
  \label{eq:disc_rotation_update}
\end{equation}
and the corresponding variation
\begin{equation}
  \delta\ten R^h=\VPTen{\mi\theta^h}\ten R^h.
  \label{eq:disc_theta}
\end{equation}
Setting \(\ten U^h:=\operatorname{exp}\VPTen{\Delta\mi\vartheta^h}\), differentiation gives, for \(\mu\in\{s,t\}\),
\[
  \ten R^{h,j+1}_{,\mu}
  =\VPTen{
      \operatorname{dexp}_{\Delta\mi\vartheta^h}
      (\Delta\mi\vartheta^h_{,\mu})
    }\ten R^{h,j+1}
   +\ten U^h\ten R^{h,j}_{,\mu},
\]
where, for \(\mi a,\mi b\in\mathbb R^3\), the differential consistent with this left multiplicative update is
\[
  \VPTen{\operatorname{dexp}_{\mi a}(\mi b)}
  :=\left.\frac{\d}{\d\varepsilon}
       \operatorname{exp}\bigl(\VPTen{\mi a+\varepsilon\mi b}\bigr)
     \right|_{\varepsilon=0}
     \operatorname{exp}\bigl(-\VPTen{\mi a}\bigr).
\]
Starting from an admissible rotation field \(\ten R^{h,0}\) and its consistent elementwise derivatives, the field and its derivatives are advanced together by these relations; derivatives at element interfaces are taken one-sided.
The discrete rotational state comprises the nodal traces and the accumulated
values of \(\ten R^h\), \(\ten R^h_{,s}\) and \(\ten R^h_{,t}\)
at the required evaluation points. The rotational states at assembly,
independent diagnostic and temporal-slice points are initialized before the
first Newton update and advanced by the same interpolated increment. At window
interfaces, the accumulated rotation and its spatial derivative are transferred
consistently.
The interior field is therefore specified by its initial state and ordered
accepted updates; it is not reconstructed from the final nodal orientations.
Group preservation and covariance under a constant frame change do not
establish independence from the Newton update path
\cite{crisfield_objectivity_1999}; the present tests do not quantify that
sensitivity.
The absolute triad follows from
\[
  \begin{aligned}
    \ten\Lambda^h&=\ten R^h\ten\Lambda_0,\\
    \ten\Lambda^h_{,s}&=\ten R^h_{,s}\ten\Lambda_0
                         +\ten R^h\ten\Lambda_{0,s},
    &\quad
    \ten\Lambda^h_{,t}&=\ten R^h_{,t}\ten\Lambda_0.
  \end{aligned}
\]
The independent momentum fields and their tests are
\begin{align}
  \mi p^h&=\sum_{B\in\mathcal I_p}M_B\mi p_B,
  &
  \mi\ell^h&=\sum_{B\in\mathcal I_p}M_B\mi\ell_B,
  \label{eq:disc_momenta}\\
  \mi\pi^h&=\sum_{B\in\mathcal I_p^{\mathrm f}}M_B\mi\pi_B,
  &
  \mi\rho^h&=\sum_{B\in\mathcal I_p^{\mathrm f}}M_B\mi\rho_B.
  \label{eq:disc_momentum_tests}
\end{align}

The reference geometry and section data are fixed as in Section~\ref{sec:cosserat_dictionary}.
For spatial interpolation \(I_h^s\) in the placement space, the two reference representations used below are
\begin{equation}
  \mi r_{0,h}:=I_h^s\mi r_0,
  \qquad
  \mi a_{0,h}:=
  \begin{cases}
    \mi D_3=\mi r_{0,s}, & \text{analytical reference},\\
    \mi r_{0,h,s}, & \text{discrete placement reference}.
  \end{cases}
  \label{eq:disc_reference_tangent}
\end{equation}
The discrete tangent is used without normalization or a change of material measure.
The second choice makes \((\mi r_{0,h},\ten I)\) strain-free while retaining the spatial geometry approximation; it is used for the coupled spring--body example.
The kinematic and constitutive quantities are
\begin{align}
  \mi\Gamma^h
  &:=(\ten R^h)\tp\mi r^h_{,s}-\mi a_{0,h}(s),
  &
  \mathcal K^h
  &:=\axl\big((\ten R^h)\tp\ten R^h_{,s}\big),
  \label{eq:disc_strains}\\
  \mi N^h
  &:=\frac{\partial\Psi}{\partial\mi\Gamma}(s,\mi\Gamma^h,\mathcal K^h),
  &
  \mi M^h
  &:=\frac{\partial\Psi}{\partial\mathcal K}(s,\mi\Gamma^h,\mathcal K^h),
  \label{eq:disc_material_resultants}\\
  \mi n^h&:=\ten R^h\mi N^h,
  &
  \mi m^h&:=\ten R^h\mi M^h,
  \label{eq:disc_spatial_resultants}\\
  \mi\omega^h
  &:=\axl\big(\ten R^h_{,t}(\ten R^h)\tp\big),
  &
  \ten J^{\mathrm{sp},h}
  &:=\ten R^h\ten J(s)(\ten R^h)\tp.
  \label{eq:disc_omega_Jsp}
\end{align}
Here \(\mathcal K^h\) is the elastic curvature change; no additional reference-curvature term is subtracted.
The local section components are \(\ten\Lambda_0\tp\mi\Gamma^h\) and \(\ten\Lambda_0\tp\mathcal K^h\), as in the continuous constitutive law.
All constitutive derivatives are taken at fixed reference data.
The pushed-forward material variations retain their form:
\begin{equation}
  \delta^\sharp\mi\gamma^h
  =\mi\eta^h_{,s}-\mi\theta^h\times\mi r^h_{,s},
  \qquad
  \delta^\sharp\mi\kappa^h
  =\mi\theta^h_{,s}.
  \label{eq:disc_pushed_variations}
\end{equation}

%------------------------------------------------------------
\subsection{Unstabilized mixed space--time form}
\label{subsec:disc_galerkin_form}

Denote the admissible trial sets by \(\mathcal Q_{\alpha,h}\), \(\alpha\in\{r,R,p,\ell\}\), and the corresponding homogeneous test spaces by \(\mathcal V_{\beta,h}\), \(\beta\in\{r,\theta,p,\ell\}\).
Introduce
\[
  \mathbf y_h:=(\mi r^h,\ten R^h,\mi p^h,\mi\ell^h),
  \qquad
  \mathbf z_h:=(\mi\eta^h,\mi\theta^h,\mi\pi^h,\mi\rho^h).
\]
The Galerkin core is the unperturbed integral form obtained from the continuously rearranged mixed system \eqref{eq:worldsheet_mixed_weak_form}; see Subsection~\ref{subsec:worldsheet_mixed_form}.
After temporal integration by parts, the two balance rows are multiplied by \(-1\), while the momentum--velocity rows retain their signs.
The continuous rearrangement is not asserted to be an algebraic row operation on the discretized mixed action: in general, \(\mi\ell^h\times\mi\theta^h\) does not belong to \(\mathcal V_{\ell,h}\), so the discarded continuous residual-proportional term need not vanish by discrete orthogonality.
The method below discretizes the rearranged balance system.
Its test space is
\[
  \mathcal Z_h:=\mathcal V_{r,h}\times\mathcal V_{\theta,h}
                  \times\mathcal V_{p,h}\times\mathcal V_{\ell,h}.
\]
The unstabilized problem is: find
\[
  \mathbf y_h\in
  \mathcal Q_{r,h}\times\mathcal Q_{R,h}
  \times\mathcal Q_{p,h}\times\mathcal Q_{\ell,h}
\]
such that
\begin{equation}
  \mathcal G_h^0(\mathbf y_h;\mathbf z_h)=0
  \qquad
  \forall\mathbf z_h\in\mathcal Z_h.
  \label{eq:disc_unstabilized_problem}
\end{equation}
The complete functional is defined by
\[
  \mathcal G_h^0
  :=\mathcal G_{h,\mathrm{vol}}^0+\mathcal G_{h,N}^0+\mathcal G_{h,t}^0,
  \qquad
  \mathcal G_{h,N}^0:=\left.\mathcal G_{h,N}^{\mathrm{PG}}\right|_{\tau_n=0},
  \quad
  \mathcal G_{h,t}^0:=\left.\mathcal G_{h,t}^{\mathrm{PG}}\right|_{\tau_n=0}.
\]
The last two terms are the spatial natural and temporal mismatch functionals in \eqref{eq:disc_stabilization_neumann_main} and \eqref{eq:disc_stabilization_flux_compact}, evaluated with unperturbed tests. Its volume contribution is
\begin{align}
  \mathcal G_{h,\mathrm{vol}}^0(\mathbf y_h;\mathbf z_h)
  :=&\sum_{a,n}\int_{\mathcal B_{an}}
  \Big[
    \mi p^h_{,t}\cdot\mi\eta^h
    +\mi n^h\cdot\mi\eta^h_{,s}
    -\mi f^h\cdot\mi\eta^h
  \nonumber\\[-0.2em]
  &\hspace{3.5em}
    +\mi\ell^h_{,t}\cdot\mi\theta^h
    +\mi m^h\cdot\mi\theta^h_{,s}
    -\big(\mi r^h_{,s}\times\mi n^h+\mi c^h\big)\cdot\mi\theta^h
  \nonumber\\[-0.2em]
  &\hspace{3.5em}
    +\big(\mi r^h_{,t}-A^{-1}\mi p^h\big)\cdot\mi\pi^h
    +\big(\mi\omega^h-(\ten J^{\mathrm{sp},h})^{-1}\mi\ell^h\big)
      \cdot\mi\rho^h
  \Big]\d s\,\d t.
  \label{eq:disc_unstabilized_form}
\end{align}
The temporal terms result from combining integration by parts with the prescribed incoming and free outgoing co-normal fluxes, as detailed in Appendix~\ref{subsec:app_temporal_completion}.
At a free terminal face the numerical outflow is the interior trace,
\begin{equation}
  \bar{\mi p}^{+}=\mi p^h|_{\Sigma_T},
  \qquad
  \bar{\mi\ell}^{+}=\mi\ell^h|_{\Sigma_T},
  \label{eq:disc_terminal_outflow_flux}
\end{equation}
so no terminal momentum is prescribed.
The upper cancellation follows from matching the traces, not from setting the outgoing momentum to zero.

%------------------------------------------------------------
\subsection{Future-directed Petrov--Galerkin stabilization}
\label{subsec:disc_stabilization}

Following Subsection~\ref{subsec:worldsheet_time_orientation}, the perturbation gives the global approximation a preferred time direction without prescribing a terminal configuration.
It is intended to reduce pre-activation response, not to prove strict domain-of-dependence causality or enforce energy conservation at every time.
The future direction used in the computations is
\[
  \beta=\partial_t.
\]
The stabilization time scale is constant on every time slab,
\begin{equation}
  \tau_n:=\theta_{\mathrm{stab}}\Delta t_n,
  \qquad
  \theta_{\mathrm{stab}}\ge0.
  \label{eq:disc_stabilization_parameter}
\end{equation}
The value \(\theta_{\mathrm{stab}}=0\) recovers the Galerkin core. All four tests are perturbed in the same future direction,
\begin{equation}
  \begin{aligned}
    \mi\eta_h^{\star,n}&:=\mi\eta_h+\tau_n\mi\eta_{h,t},
    &\qquad
    \mi\theta_h^{\star,n}&:=\mi\theta_h+\tau_n\mi\theta_{h,t},
    \\
    \mi\pi_h^{\star,n}&:=\mi\pi_h+\tau_n\mi\pi_{h,t},
    &
    \mi\rho_h^{\star,n}&:=\mi\rho_h+\tau_n\mi\rho_{h,t}.
  \end{aligned}
  \label{eq:disc_stabilized_test}
\end{equation}
Here \(\Delta t_n\) is the duration of a temporal element, not the duration of a solution window containing several such elements.
For a uniform partition of a window of length \(H\) into \(N_t\) temporal elements, \(\tau=\theta_{\mathrm{stab}}H/N_t\).

The starred fields are elementwise Petrov--Galerkin images of the original tests in \(\mathcal Z_h\); they need not belong to that same continuous space.
On an aligned element, inserting them into the integral form gives
\begin{align}
  \mathcal G_{h,\mathrm{vol}}^{\mathrm{PG}}
  :=&\sum_{a,n}\int_{\mathcal B_{an}}
  \Big[
    \mi p^h_{,t}\cdot\mi\eta_h^{\star,n}
    +\mi n^h\cdot(\mi\eta_h^{\star,n})_{,s}
    -\mi f^h\cdot\mi\eta_h^{\star,n}
  \nonumber\\[-0.2em]
  &\hspace{3.5em}
    +\mi\ell^h_{,t}\cdot\mi\theta_h^{\star,n}
    +\mi m^h\cdot(\mi\theta_h^{\star,n})_{,s}
    -\big(\mi r^h_{,s}\times\mi n^h+\mi c^h\big)
      \cdot\mi\theta_h^{\star,n}
  \nonumber\\[-0.2em]
  &\hspace{3.5em}
    +\big(\mi r^h_{,t}-A^{-1}\mi p^h\big)\cdot\mi\pi_h^{\star,n}
    +\big(\mi\omega^h-(\ten J^{\mathrm{sp},h})^{-1}\mi\ell^h\big)
      \cdot\mi\rho_h^{\star,n}
  \Big]\d s\,\d t.
  \label{eq:disc_stabilized_volume_form}
\end{align}
Equivalently,
\begin{equation}
  \mathcal G_{h,\mathrm{vol}}^{\mathrm{PG}}
  =\mathcal G_{h,\mathrm{vol}}^0+\mathcal S_h^{\mathrm{vol,w}},
\end{equation}
where the weak-form perturbation is
\begin{align}
  \mathcal S_h^{\mathrm{vol,w}}
  :=&\sum_{a,n}\int_{\mathcal B_{an}}\tau_n
  \Big[
    \mi p^h_{,t}\cdot\mi\eta^h_{,t}
    +\mi n^h\cdot\mi\eta^h_{,st}
    -\mi f^h\cdot\mi\eta^h_{,t}
  \nonumber\\[-0.2em]
  &\hspace{3.5em}
    +\mi\ell^h_{,t}\cdot\mi\theta^h_{,t}
    +\mi m^h\cdot\mi\theta^h_{,st}
    -\big(\mi r^h_{,s}\times\mi n^h+\mi c^h\big)
      \cdot\mi\theta^h_{,t}
  \nonumber\\[-0.2em]
  &\hspace{3.5em}
    +\big(\mi r^h_{,t}-A^{-1}\mi p^h\big)\cdot\mi\pi^h_{,t}
    +\big(\mi\omega^h-(\ten J^{\mathrm{sp},h})^{-1}\mi\ell^h\big)
      \cdot\mi\rho^h_{,t}
  \Big]\d s\,\d t.
  \label{eq:disc_stabilization_volume}
\end{align}
This is the implemented complete Petrov--Galerkin volume form.
For the spatial natural terms we use the material pullback of the geometric co-normal work in \eqref{eq:worldsheet_natural_virtual_work}.
On an aligned beam end, the pullbacks of \(\bar{\mathfrak t}_r^h\,\d\ell\) and \(\bar{\mathfrak t}_\theta^h\,\d\ell\) are \(\bar{\mi t}^h\,\d t\) and \(\bar{\mi m}^h\,\d t\), respectively, by \eqref{eq:worldsheet_conormal_pullback}.
The superscript \(h\) records the chosen representation of the prescribed data.
Using the same Petrov--Galerkin images, the natural functional is therefore
\begin{align}
  \mathcal G_{h,N}^{\mathrm{PG}}
  :=&-\int_{\mathcal T}
    \sum_{s_e\in\Gamma_r^N}
    \bar{\mi t}^h(s_e,t)\cdot\mi\eta_h^{\star}(s_e,t)\,\d t
    \nonumber\\
    &-\int_{\mathcal T}
    \sum_{s_e\in\Gamma_R^N}
    \bar{\mi m}^h(s_e,t)\cdot\mi\theta_h^{\star}(s_e,t)\,\d t.
  \label{eq:disc_stabilization_neumann_main}
\end{align}
The temporal mismatch terms use the one-sided starred traces from the first and last time slabs, respectively:
\begin{align}
  \mathcal G_{h,t}^{\mathrm{PG}}
  :=&\int_{\Sigma_T}
  \Big[
    (\bar{\mi p}^{+}-\mi p^h)\cdot\mi\eta_h^{\star}
    +(\bar{\mi\ell}^{+}-\mi\ell^h)\cdot\mi\theta_h^{\star}
  \Big]\d s
  \nonumber\\
  &+\int_{\Sigma_0}
  \Big[
    (\mi p^h-\bar{\mi p}^{-})\cdot\mi\eta_h^{\star}
    +(\mi\ell^h-\bar{\mi\ell}^{-})\cdot\mi\theta_h^{\star}
  \Big]\d s.
  \label{eq:disc_stabilization_flux_compact}
\end{align}
Under the outflow choice \eqref{eq:disc_terminal_outflow_flux}, the upper term vanishes because the physical and numerical momentum traces coincide.
The lower mismatch vanishes because the incoming momentum traces are prescribed strongly.
Both terms are displayed before these trace identities are applied. Although the original tests vanish on \(\Sigma_0\), their temporal derivatives, and hence their starred traces, need not vanish there; this test condition alone cannot justify discarding the temporal work.

No interior facet parameter is required for the aligned continuous tensor-product spaces defined above.
Here this means a parameter controlling the numerical co-normal transfer between neighboring elements, such as \(\alpha_F\) in \eqref{eq:app_flux_family}; it is distinct from \(\theta_{\mathrm{stab}}\).
On a spatial interface inside one time slab, \(\partial_t\mathbf z_h\) is the tangential derivative of the common trace.
Since \(\tau_n\) is the same on both spatially adjacent elements, the corresponding starred traces agree there.
On a constant-time interface \(F\), the starred traces can differ, even though the original tests and the momenta are continuous.
Choosing the common numerical traces
\(\widehat{\mi p}_F=\mi p^{h,-}=\mi p^{h,+}\) and
\(\widehat{\mi\ell}_F=\mi\ell^{h,-}=\mi\ell^{h,+}\)
makes each one-sided temporal mismatch vanish:
\[
  (\widehat{\mi p}_F-\mi p^{h,\pm})\cdot\mi\eta_h^{\star,\pm}=0,
  \qquad
  (\widehat{\mi\ell}_F-\mi\ell^{h,\pm})\cdot\mi\theta_h^{\star,\pm}=0.
\]
Thus the temporal integration-by-parts terms are canceled by their matching local flux terms; continuity of the momenta alone would not cancel raw products with unequal starred test traces.
No additional temporal interface term remains in the displayed formulation with derivatives on the momenta.
The non-aligned case, in which \(D_\beta\mathbf z_h\) has two genuinely different traces, is treated separately in Appendix~\ref{subsec:app_non_aligned_facets}.

The complete stabilized problem is
\begin{equation}
  \boxed{
  \mathcal G_h^{\mathrm{PG}}(\mathbf y_h;\mathbf z_h)
  :=\mathcal G_{h,\mathrm{vol}}^{\mathrm{PG}}
    +\mathcal G_{h,N}^{\mathrm{PG}}
    +\mathcal G_{h,t}^{\mathrm{PG}}
  =0
  \quad\forall\mathbf z_h\in\mathcal Z_h.}
  \label{eq:disc_stabilized_problem}
\end{equation}
For later balance identities we write
\begin{equation}
  \mathcal S_h^{\mathrm{PG}}
  :=\mathcal G_h^{\mathrm{PG}}-\mathcal G_h^0.
  \label{eq:disc_complete_pg_perturbation}
\end{equation}

Element residuals and balance identities are detailed in Appendices~\ref{subsec:app_complete_pg_decomposition} and~\ref{subsec:app_balance_defects}.

%------------------------------------------------------------
\subsection{Solution windows}
\label{subsec:disc_newton}

All temporal elements within a solution window are solved together.
For successive windows, the accepted terminal configuration and momenta supply the strongly prescribed incoming data of the next window, using the same spatial discretization and fixed reference data.
The rotation trace and its spatial derivative are transferred consistently, and external loads retain their physical time argument.
The window partition is part of the temporal discretization and is specified separately from the element sizes \(\Delta t_n\).

The nonlinear system is solved by Newton's method with a consistent tangent and the multiplicative rotation update \eqref{eq:disc_rotation_update}.
The stopping criterion uses the absolute Euclidean norm of the active assembled residual; the tolerances are specified with each numerical study.

%------------------------------------------------------------
\subsection{Balances and monitored defects}
\label{subsec:disc_balances_defects}

The stabilized method solves the weak momentum equations, whereas the independent momentum fields need not satisfy the temporal constitutive relations pointwise on a finite mesh.
We therefore monitor
\begin{equation}
  \mi d_p^h:=\mi r^h_{,t}-A^{-1}\mi p^h,
  \qquad
  \mi d_\ell^h:=\mi\omega^h-(\ten J^{\mathrm{sp},h})^{-1}\mi\ell^h.
  \label{eq:disc_monitored_mixed_defects}
\end{equation}
For nonzero normalizing scales, the corresponding ideal relative measures are
\begin{align}
  \delta_p^h
  &:=\frac{\|\mi d_p^h\|_{L^2(\mathcal B_0)}}
  {\left[\tfrac12\left(
    \|\mi r^h_{,t}\|_{L^2(\mathcal B_0)}^2
    +\|A^{-1}\mi p^h\|_{L^2(\mathcal B_0)}^2
  \right)\right]^{1/2}},
  \\
  \delta_\ell^h
  &:=\frac{\|\mi d_\ell^h\|_{L^2(\mathcal B_0)}}
  {\left[\tfrac12\left(
    \|\mi\omega^h\|_{L^2(\mathcal B_0)}^2
    +\|(\ten J^{\mathrm{sp},h})^{-1}\mi\ell^h\|_{L^2(\mathcal B_0)}^2
  \right)\right]^{1/2}}.
\end{align}
If a denominator is zero, both compared rates and the associated defect vanish in \(L^2\); the corresponding relative measure is then defined as zero.
The reported numerical values replace the squared \(L^2\)-norms by sums with the chosen quadrature weights and material element measures.
For an absolute defect square \(D_{\mathrm Q}^2\) and the corresponding symmetric scale square \(S_{\mathrm Q}^2\), the implementation evaluates
\begin{equation}
  \delta_{\mathrm Q}
  :=\left(\frac{D_{\mathrm Q}^2}
                 {\max\{S_{\mathrm Q}^2,\varepsilon_{\mathrm{fp}}\}}\right)^{1/2},
  \label{eq:disc_numerical_defect_normalization}
\end{equation}
separately for the translational and rotational relation.
Here \(\varepsilon_{\mathrm{fp}}\) is machine epsilon applied as an absolute safeguard to the numerical squared scale in the chosen units.
When that safeguard is active, the value is not interpreted as a scale-independent relative error; the absolute defect norms are reported as well, particularly for nearly quiescent states.
The assembled residuals are driven to the nonlinear-solver tolerance for the chosen quadrature; reintegration with a different rule is a separate diagnostic.
Thus the strong defects are compatibility and approximation measures, not residuals of unsolved discrete equations.

A constant translational test is unchanged by the Petrov--Galerkin map whenever it is admissible. This observation alone does not establish a balance on an arbitrary time subinterval, particularly when the initial trace or support conditions exclude that test.
Appendix~\ref{subsec:app_balance_defects} states the associated balance diagnostics with these restrictions made explicit.
Energy on fixed-time sections is monitored using the momentum-based Hamiltonian in \eqref{eq:disc_total_energy} and the associated work remainder in \eqref{eq:disc_energy_balance_defect}.
Because the momentum-derived rates need not equal the configuration rates pointwise, this diagnostic can contain a power mismatch in addition to spatial, temporal, quadrature and Petrov--Galerkin contributions.
It is an approximation diagnostic; no pointwise-in-time conservation condition is imposed.

% Version 8 (27 October 2026): self-contained condensed numerical examples.
% Completed M2 controls update the two-second spring illustration.
% The short ST/MP convergence study remains at T=0.025 s.
\section{Numerical examples}
\label{sec:numerical_examples}

The tests separate verification of the space--time balance and boundary
structure from approximation accuracy. They cover temporal outflow,
smooth-solution convergence, spatial shear control, conventional time
marching, delayed loading, covariance, finite rotations and non-aligned
meshes, followed by a spring--rigid-body application. The spring study uses a separately archived
campaign; its relevant mechanical formulation and output definitions were
checked across the candidate and reference source versions.

\paragraph{Common setup and diagnostics.}
Unless stated otherwise, placement and both momentum fields use continuous
biquadratic tensor-product bases, with the same nominal orders for
rotational increments and virtual spins. The finite rotation is the
recursively updated group-valued field of
Subsection~\ref{subsec:disc_fields}. We write \([N_s,N_t]\) for the
spatial and temporal element counts and use \(\beta=\partial_t\),
\(\tau_n=\theta_{\mathrm{stab}}\Delta t_n\). On a uniform quadratic mesh,
the temporal nodal spacing is \(\delta t_{\mathrm{node}}=t_{\mathrm{end}}/(2N_t)\),
whereas stabilization uses the element width
\(\Delta t_n=t_{\mathrm{end}}/N_t\).
Table~\ref{tab:common-numerical-data} gives the straight-beam data; the
spring data are specified in Subsection~\ref{subsec:num_spring_rigid_body}.
The stored energy is
\(\Psi=\tfrac12\mi\Gamma\cdot\ten C_\Gamma\mi\Gamma
+\tfrac12\K\cdot\ten C_K\K\).
Full transverse-shear integration is the default. Selective reduced
integration (SRI) is identified wherever used.

\begin{table}[tbp]
\centering
\small
\caption{Common straight-beam data in a consistent system of units.
The cross-section area \(A_{\mathrm{cs}}\) is distinct from the
translational line inertia \(A(s)=m_0\). Departures are stated with each test.}
\label{tab:common-numerical-data}
\begin{tabular}{ll}
\toprule
quantity & value\\
\midrule
reference & \(\mi r_0=s\mi e_3,\quad\boldsymbol\Lambda_0=\ten I,\quad\mi a_{0,h}=\mi D_3=\mi e_3\)\\
length and time scaling & \(L=1,\quad c_\tau=1\)\\
material & \(E=10^4,\quad G=4\times10^3,\quad\rho=1\)\\
section and line mass & \(A_{\mathrm{cs}}=10^{-2},\quad m_0=\rho A_{\mathrm{cs}}=10^{-2}\)\\
second moments & \(I_1=I_2=10^{-6}\)\\
\(\ten C_\Gamma\) & \(\operatorname{diag}(GA_{\mathrm{cs}},GA_{\mathrm{cs}},EA_{\mathrm{cs}})\)\\
\(\ten C_K\) & \(\operatorname{diag}(EI_1,EI_2,G(I_1+I_2))\)\\
\(\ten J\) & \(\operatorname{diag}(\rho I_1,\rho I_2,\rho(I_1+I_2))\)\\
main Q2/Q2 parameter & \(\theta_{\mathrm{stab}}=0.05\)\\
Q2 assembly / diagnostic rule & \(3\times3\) / positive \(4\times4\)\\
P2 assembly rule & 13-point triangle rule, exactness degree 7\\
P2 diagnostic rule & positive 28-point triangle rule, exactness degree 11\\
linear solver & direct sparse solve\\
\bottomrule
\end{tabular}
\end{table}

Assembly quadrature forms the algebraic residual and tangent; diagnostic
quadrature independently integrates field errors and compatibility defects.
A positive rule has positive weights; this does not assert a vanishing
error or residual.
For the manufactured solution, \(\|\cdot\|_Q\) denotes the quadrature
evaluation of the material space--time \(L^2\)-norm with measure
\(\d s\,\d t\). The relative errors are
\begin{equation}
 \begin{aligned}
 e_r&=\frac{\|\mi r^h-\mi r^{\mathrm{ex}}\|_Q}
                 {\|\mi r^{\mathrm{ex}}-\mi r_0\|_Q},
 &e_R&=\frac{\|\operatorname{Log}(\ten R^h\ten R^{\mathrm{ex}\tp})\|_Q}
                 {\|\operatorname{Log}(\ten R^{\mathrm{ex}})\|_Q},\\
 e_z&=\frac{\|\mi z^h-\mi z^{\mathrm{ex}}\|_Q}
                 {\|\mi z^{\mathrm{ex}}\|_Q},
 &\mi z&\in\{\mi p,\mi\ell,\mi n,\mi m\}.
 \end{aligned}
 \label{eq:num_relative_field_errors}
\end{equation}
Here \(\operatorname{Log}(\ten Q)=\operatorname{axl}(\log\ten Q)\)
is the principal rotation vector, and \(\mi n,\mi m\) are spatial
resultants. Placement is normalized by displacement rather than absolute
position. Field-error denominators use
\(\max(\|\text{reference}\|_Q,\varepsilon_{\mathrm{fp}})\);
the rate-based compatibility defects instead use the squared-scale
safeguard of \eqref{eq:disc_numerical_defect_normalization}.
Absolute errors, reference scales and safeguard activation are retained.
An unscaled algebraic residual is the Euclidean norm of the active
coefficient vector, not a mesh-independent physical error norm. The
compatibility rows \(\mathcal R_p,\mathcal R_\ell\) act on free
momentum tests and differ from both the mechanical balance rows
\(\mathcal R_r,\mathcal R_\theta\) and the strong defects.
Straight-beam Newton tolerances are \(10^{-8}\) for manufactured-solution
and simplex sequences, \(10^{-11}\) for the outflow audit,
\(10^{-9}\) for the multimode free-bending comparison, and \(10^{-10}\)
for delayed loading and covariance. Reduced-amplitude manufactured
controls use \(10^{-11}\) when the amplitudes are multiplied by \(10^{-3}\).
Physical initial conditions prescribe traces of all four fields;
case-specific initial iterates away from that face are numerical starting
values, not additional physical conditions.

The delayed-load and spring examples use the smooth ramp
\[
q(\xi)=
\begin{cases}
0, & \xi\le 0,\\[1mm]
10\xi^3-15\xi^4+6\xi^5, & 0<\xi<1,\\[1mm]
1, & \xi\ge 1.
\end{cases}
\]
Its argument and load magnitude are specified with each example.

\subsection{Consistency of the free temporal boundary}
\label{subsec:num_temporal_outflow}

The end of the computational time interval imposes no physical
constraint on a freely moving beam.
This test checks that the implemented space--time equations admit
a known motion with nonzero terminal momenta without prescribing
terminal data.

We consider a traction-free straight beam undergoing uniform
translation and steady spin about its centerline.
With \(t_{\mathrm{end}}=0.30\), \(I_2=1.7\times10^{-6}\),
\(\mi v_0=(0.12,-0.07,0.05)\tp\), and \(\Omega=1.75\),
the exact motion is
\[
 \begin{aligned}
 \mi r^{\mathrm{ex}}&=s\mi e_3+t\mi v_0,
 &\ten R^{\mathrm{ex}}&=
   \exp(\VPTen{\Omega t\mi e_3}),\\
 \mi p^{\mathrm{ex}}&=m_0\mi v_0,
 &\mi\ell^{\mathrm{ex}}&=
   \rho(I_1+I_2)\Omega\mi e_3.
 \end{aligned}
\]
Body loads and stress resultants vanish.
The spatially integrated momentum magnitudes remain constant:
\(\|\int_{\mathfrak I}\mi p^{\mathrm{ex}}\,\d s\|
 \simeq1.476\times10^{-3}\) and
\(\|\int_{\mathfrak I}\mi\ell^{\mathrm{ex}}\,\d s\|
 =4.725\times10^{-6}\).
The latter is the intrinsic angular momentum, excluding the
orbital contribution.

The exact history is inserted into all four discrete fields
on the mesh \([8,16]\).
Only the initial traces are fixed; the terminal face uses the
interior momentum traces as its outgoing fluxes according to
\eqref{eq:disc_terminal_outflow_flux}.
We evaluate the assembled residual at this known state without
performing a Newton solve.
The recorded active algebraic residual norm is
\(3.280\times10^{-14}\).

To identify the role of the terminal momentum contribution,
we evaluate the residual increment associated with omitting
only that contribution while keeping all fields unchanged.
In the rearranged balance equations, this increment is
\[
 \Delta\mi{\mathcal R}_{r,A}^{T}
 =-\int_{\Sigma_T}N_A^\star\mi p^h\,\d s,
 \qquad
 \Delta\mi{\mathcal R}_{\theta,A}^{T}
 =-\int_{\Sigma_T}N_A^\star\mi\ell^h\,\d s.
\]
Its Euclidean norms over the active translational and rotational
coefficients are \(4.517\times10^{-4}\) and
\(1.446\times10^{-6}\), respectively.
These are algebraic residual increments, distinct from the
physical momentum magnitudes given above.

The check confirms that the implemented residual is consistent
with this free motion, whereas omission of the outgoing
contribution leaves a nonzero residual at the same state.
In the continuous variational statement, such an omission
with a free terminal configuration would impose zero terminal
momenta; see Subsection~\ref{subsec:worldsheet_weak_fluxes}.
The nonzero terminal momenta make the present motion suitable
for detecting this omission, which would be invisible at the
exact solution of Subsection~\ref{subsec:num_manufactured_q2},
where both terminal momenta vanish.

\subsection{Smooth-solution convergence and spatial shear control}
\label{subsec:num_manufactured_q2}

Let \(t_{\mathrm{end}}=0.25\), \(k=\pi/(2L)\),
\(\xi=t/t_{\mathrm{end}}\), and \(g(t)=64\xi^3(1-\xi)^3\).
The manufactured motion is
\begin{equation}
 \mi r^{\mathrm{ex}}=
 \begin{bmatrix}a\sin(ks)g(t)\\0\\s\end{bmatrix},
 \qquad
 \ten R^{\mathrm{ex}}=
 \exp\!\left(\VPTen{b\sin(ks)g(t)\mi e_2}\right),
 \label{eq:num_manufactured_solution}
\end{equation}
with \(a=5\times10^{-4}\), \(b=10^{-3}\),
\(\mi p^{\mathrm{ex}}=m_0\mi r^{\mathrm{ex}}_{,t}\) and
\(\mi\ell^{\mathrm{ex}}=\ten J^{\mathrm{sp,ex}}\mi\omega^{\mathrm{ex}}\).
Placement and orientation are fixed at \(s=0\); distributed loads and
right-end natural data follow by substitution. The initial traces are
\(\mi r=\mi r_0\), \(\ten R=\ten I\),
\(\mi p=\mi\ell=\mi0\), and the convergence runs start from the rest
history. An interpolated exact history is used only in the separately
identified initialization and linearization audit. Since
\(g=g_{,t}=g_{,tt}=0\) at both temporal endpoints, this test has neither
a delayed onset nor a nonzero terminal-momentum layer.

\begin{figure}[tbp]
 \centering
 \includegraphics[width=0.82\textwidth]{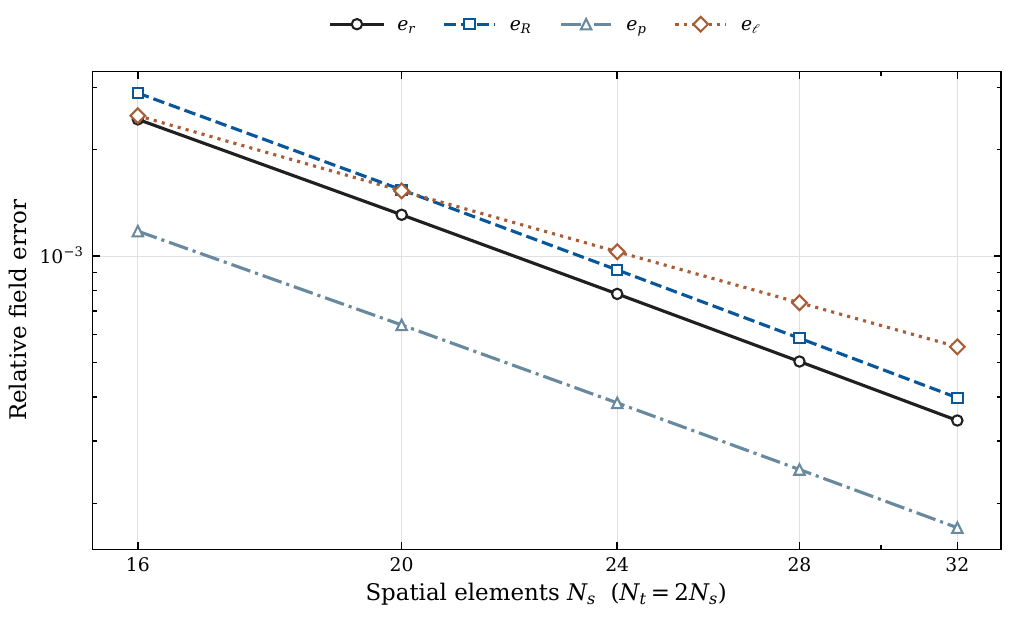}
 \caption{Primary-field convergence for the smooth manufactured solution
 under proportional refinement with \(N_t=2N_s\), evaluated by the positive
 \(4\times4\) diagnostic rule.}
 \label{fig:num_q2_manufactured_primary}
\end{figure}

We refine the space--time mesh using
\(N_s=16,20,24,28,32\) elements along the beam and
\(N_t=2N_s\) elements in time.
All monitored errors decrease monotonically along this sequence.
Figure~\ref{fig:num_q2_manufactured_primary} shows the corresponding
primary-field errors.
The final-step rates are approximately \(2.88,2.90,2.84,2.15,1.87,2.85\)
for \(\mi r,\ten R,\mi p,\mi\ell,\mi n,\mi m\), respectively;
the strong mixed defects have rates close to two. These are observed
rates under proportional space--time refinement, not a general
optimal-order result. The maximum relative difference between field
errors evaluated with the assembly and positive diagnostic quadrature
rules is \(1.644\times10^{-3}\), using the diagnostic-rule error for
normalization. This comparison does not bound the quadrature error.

\paragraph{Spatial shear control.}
The planar spatial tangent is extracted from the native element at the
straight state using fields constant in time. Linearized end-moment and
end-force tests distinguish shear locking from temporal error. With
\(S=GA_{\mathrm{cs}}\), \(B=EI_2\), \(\Lambda=SL^2/B\), the exact
end-moment solution is \(\varphi(s)=Ms/B\), \(w(s)=Ms^2/(2B)\).
For fully integrated Q1 elements, the tip ratio is \((1+\Lambda/(12N_s^2))^{-1}\); selective
reduction recovers the exact nodal tip value, but not the quadratic
deflection between nodes. For the degree-\(p\) controls, SRI reduces only spatial transverse-shear
quadrature from \(p+1\) to \(p\) points, retaining full integration of
the other terms and unchanged temporal quadrature. The end-force
reference is \(w_{\mathrm{ex}}(L)=FL^3/(3B)+FL/S\). For two Q2 elements
at \(\Lambda=4000\), the normalized tip response is
\(0.941082\) with full integration and \(1.000000\) with SRI.
Both Q2 variants also reproduce the end-moment tip response.
Agreement with the analytical tip responses provides a numerical
indicator of locking-free behaviour of the SRI formulation in these
bending tests.

\subsection{Comparison with an over-resolved implicit-midpoint reference}
\label{subsec:num_midpoint_reference}

\paragraph{Multimode free bending.}
A clamped, initially undeformed beam vibrates freely over
\(t_{\mathrm{end}}=1.5\), using 32 quadratic spatial elements and full
shear integration. With \(\xi=s/L\), the initial rates are
\[
 \mi v(s,0)=0.05\xi^2(3-2\xi)\mi e_1,
 \qquad
 \mi\omega(s,0)=\frac{0.30}{L}\xi(1-\xi)\mi e_2.
\]
The initial momenta are \(\mi p=m_0\mi v\),
\(\mi\ell=\ten J\mi\omega\). The global initial iterate extends
undeformed placement, identity rotation and these momentum profiles
across time, fixing only the prescribed boundary and initial coefficients.
The midpoint calculations in this subsection use componentwise spatial
interpolation of placement and directors, with interpolated Lagrange
multiplier fields weakly enforcing director orthonormality. A consistent mass
matrix is assembled in these Cartesian variables, and the implicit
midpoint equations are solved with additive Newton updates. The
space--time rotation field instead follows
Subsection~\ref{subsec:disc_fields}.

The 131072-step midpoint reference differs from the 65536-step run by
\(2.645286\times10^{-7}\) in relative tip-history norm. On every candidate
grid, this change is below one percent of its candidate error; the largest
ratio is \(0.4346\%\). This is a resolution-sensitivity check, not a
rigorous reference-error bound. The material and section coefficients
agree to a relative mismatch of \(8.67\times10^{-16}\), while the two
methods retain different nonlinear rotation and inertia operators.

\begin{figure}[tbp]
 \centering
 \includegraphics[width=0.84\textwidth]{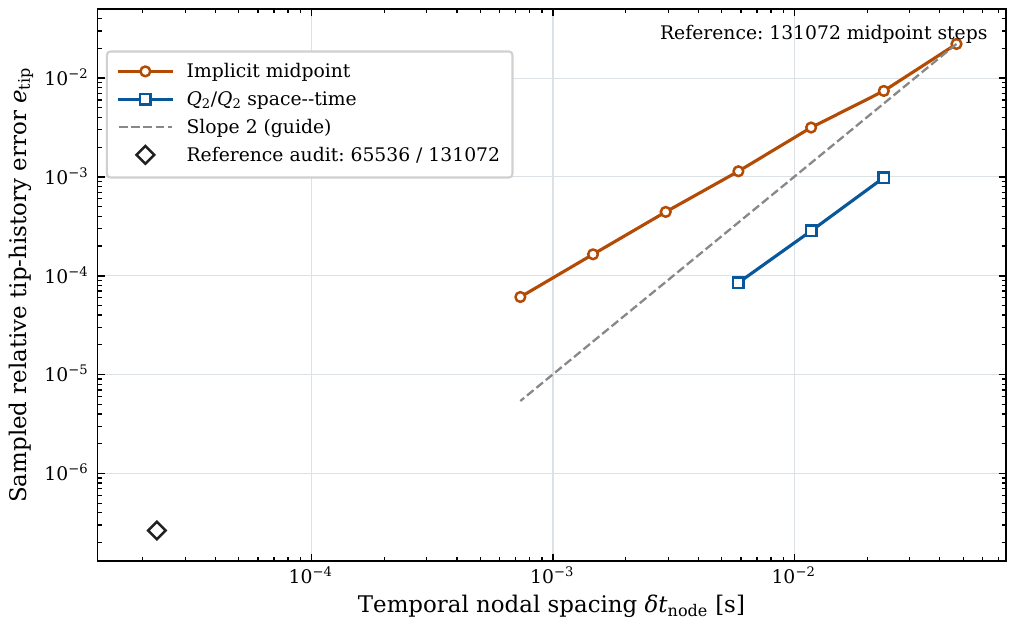}
 \caption{Sampled relative tip-history errors in the candidate-node
 trapezoidal norm against the 131072-step midpoint reference. The isolated
 marker is the 65536/131072 reference self-audit; the slope-two line is
 a visual guide.}
 \label{fig:num_midpoint_convergence}
\end{figure}

We compare the transverse tip-displacement histories at the
candidate nodal times \(t_0<\cdots<t_M\).
For these samples, define the composite trapezoidal rule
\[
 \mathcal Q_{\mathrm{trap}}[f]
 :=
 \sum_{i=0}^{M-1}
 \frac{t_{i+1}-t_i}{2}
 \bigl[f(t_i)+f(t_{i+1})\bigr].
\]
The sampled relative tip-history error is
\[
 e_{\mathrm{tip}}
 =
 \left[
 \frac{
   \mathcal Q_{\mathrm{trap}}
   \bigl[(u_{x,h}-u_{x,\mathrm{ref}})^2\bigr]
 }{
   \mathcal Q_{\mathrm{trap}}
   \bigl[u_{x,\mathrm{ref}}^2\bigr]
 }
 \right]^{1/2}.
\]
The sample times are candidate nodes, all contained in the fine reference
grid; this is not a space--time field norm. The Q2/Q2 meshes
\([32,32]\), \([32,64]\), \([32,128]\), with
\(\theta_{\mathrm{stab}}=0.05\), match the nodal spacings of 64, 128 and
256 midpoint steps. Their smaller errors in
Figure~\ref{fig:num_midpoint_convergence} are an accuracy observation at
matching nodal spacings, not an equal-cost comparison or an assertion
of identical nonlinear spatial operators.

The multimode midpoint curve does not show a uniformly quadratic slope.
An independent linearized planar Q2 modal calculation with the same
spatial resolution and excitation reproduces the seven native midpoint
errors within approximately \(0.04301\%\). Its frequency map is
\[
 \widetilde\omega=\frac{2}{\Delta t_{\mathrm{MP}}}
 \arctan\!\left(\frac{\omega\Delta t_{\mathrm{MP}}}{2}\right).
\]
Different modes accumulate different phase errors in the tip history.
Coincident candidate and reference nodes exclude between-node reference
interpolation as the cause of this curve. These checks support a
pre-asymptotic multimode interpretation; they do not prove the asymptotic
order of the nonlinear comparison. The modal reproduction uses the final
reference resolution and the repeated native data. The separate
single-mode control below tests temporal order without replacing this
multimode result.

\paragraph{Mechanical energy at fixed times.}
The momentum-based beam energy is a spatial integral at fixed time,
\begin{equation}
 E_h(t)=\int_{\mathfrak I}\left[
 \frac{\mi p^h\cdot\mi p^h}{2m_0}
 +\frac12\mi\ell^h\cdot(\ten J^{\mathrm{sp},h})^{-1}\mi\ell^h
 +\Psi(\mi\Gamma^h,\K^h)\right]\,\d s.
 \label{eq:num_section_energy}
\end{equation}
For \(E_0=E_h(0)\) and the set \(\mathcal T_{\mathrm{node}}\) of all
quadratic temporal nodes, including interior nodes, define
\(d_{E,\mathrm{node}}=\max_{t\in\mathcal T_{\mathrm{node}}}
|E_h(t)-E_0|/E_0\). On the three Q2/Q2 meshes above,
\(E_0\simeq4.644\times10^{-6}\), and the recorded maxima are
\(1.714\times10^{-3}\), \(4.373\times10^{-4}\) and
\(1.145\times10^{-4}\). They are finite-sample values, not
continuous-time maxima. The fixed support does no physical work.
Changes in \(E_h\) differ from the residual-work remainder in
\eqref{eq:disc_energy_balance_defect}, which also retains
momentum-derived boundary and load power.

\paragraph{Single-mode temporal control.}
An independent control fixes the native spatial operator linearized about
the straight state, with eight quadratic elements. For a constrained
eigenmode \(\ten K_s\mi\phi=\omega_h^2\ten M_s\mi\phi\),
\(\mi\phi\tp\ten M_s\mi\phi=1\), the exact temporal reference is
\[
 \mi q_{\mathrm{ref}}(t)=a_m\sin(\omega_h t)\mi\phi,
 \qquad
 \mi v_{\mathrm{ref}}(t)=a_m\omega_h\cos(\omega_h t)\mi\phi.
\]
The initial state is \(\mi q(0)=0\),
\(\mi v(0)=a_m\omega_h\mi\phi\); the reference amplitude gives a
maximum nodal translation of \(10^{-3}L\). Runs cover two periods,
\(T=4\pi/\omega_h\).
The mode is not imposed as a restriction on the subsequent space--time
solution. This reference is exact for the linearized spatial system,
not the continuum beam or arbitrary finite motions.

The midpoint metric uses the modal coordinate
\(a_h(t_i)=\mi\phi\tp\ten M_s\mi q_h(t_i)\) at step ends.
The space--time metric instead uses the full spatial coordinate vector:
\[
 \begin{aligned}
 e_{\mathrm{mode,node}}^2
 &=\frac{\mathcal Q_{\mathrm{trap}}
        [(a_h(t_i)-a_m\sin(\omega_h t_i))^2]}
        {\mathcal Q_{\mathrm{trap}}[a_m^2\sin^2(\omega_h t_i)]},\\
 e_{q,M}^2
 &=\frac{\int_0^T\|\mi q_h-\mi q_{\mathrm{ref}}\|_{M_s}^2\,\d t}
        {\int_0^T\|\mi q_{\mathrm{ref}}\|_{M_s}^2\,\d t},
 \qquad \|\mi z\|_{M_s}^2=\mi z\tp\ten M_s\mi z.
 \end{aligned}
\]
The latter evaluates native temporal polynomials with 12 Gauss points
per element, checked against 20 points. For the first axial mode with full integration,
midpoint errors at 16, 32, 64 and 128 steps are
\(0.3316755\), \(0.09074412\), \(0.02323279\) and
\(0.005843185\), giving rates \(1.870\), \(1.966\) and \(1.991\).
The endpoints also agree with the closed modal phase
\(2n\arctan(\omega_h\Delta t_{\mathrm{MP}}/2)\).
For quadratic space--time elements at
\(\theta_{\mathrm{stab}}=0.05\),
\(e_{q,M}\) decreases from \(8.390535\times10^{-3}\)
at \(N_t=16\) through \(1.479659\times10^{-3}\)
at \(N_t=32\) to \(2.176037\times10^{-4}\)
at \(N_t=64\), corresponding to observed temporal
convergence rates of \(2.503\) and \(2.765\).

%============================================================
\subsection{Future-directed stabilization under a delayed end load}
\label{subsec:num_q2_causality}

A clamped beam starts from the undeformed rest state
\((\mi r_0,\ten I,\mi0,\mi0)\), with the same history as the Newton
initial iterate. Only the four initial traces and the left-end placement
and orientation are prescribed. Distributed loads vanish, and the
right-end data are
\[
 \bar{\mi t}(L,t)=2\times10^{-4}
 q\!\left(\frac{t-t_{\mathrm{on}}}{t_{\mathrm{rise}}}\right)\mi e_1,
 \qquad \bar{\mi m}(L,t)=\mi0,
\]
where \(t_{\mathrm{end}}=0.25\), \(t_{\mathrm{on}}=0.125\), and
\(t_{\mathrm{rise}}=t_{\mathrm{end}}/16\). The exact initial-value
solution remains at rest before activation. A computed response in that
interval therefore measures a pre-activation component of the global
approximation. The aligned \(Q_2/Q_2\) study uses
\(\theta_{\mathrm{stab}}\in\{0,0.01,0.025,0.05,0.10\}\) on
\([8,16]\), \([12,24]\), and \([16,32]\), with a fine
\([32,64]\), \(\theta_{\mathrm{stab}}=0.05\) working reference.

For \(z\in\{u_{x,\mathrm{tip}},\|\mi p_{\mathrm{tip}}\|,
\|\mi\ell_{\mathrm{tip}}\|\}\), define the fixed reference scale and
pre-response indicator by
\[
 s_z=\max_{t_j^{\mathrm{ref}}>t_{\mathrm{on}}}
       |z_{\mathrm{ref}}(t_j^{\mathrm{ref}})|,
 \qquad
 P_z=\frac{\max_{t_i<t_{\mathrm{on}}}|z_h(t_i)|}{s_z}.
\]
The same scale is used for every candidate; maxima use stored temporal
nodes, excluding onset nodes within a roundoff tolerance. The post-onset
error instead uses the full tip-displacement vector:
\[
 e_{\mi u,\mathrm{post}}^2=
 \frac{\mathcal Q_{\mathrm{trap}}
   \|\mi u_{\mathrm{tip},h}(t_i)-\mi u_{\mathrm{tip,ref}}(t_i)\|^2}
 {\mathcal Q_{\mathrm{trap}}
   \|\mi u_{\mathrm{tip,ref}}(t_i)\|^2},
\]
with the reference evaluated at candidate nodes by shape-preserving
piecewise cubic interpolation.

\begin{figure}[tbp]
  \centering
  \includegraphics[width=0.78\textwidth]{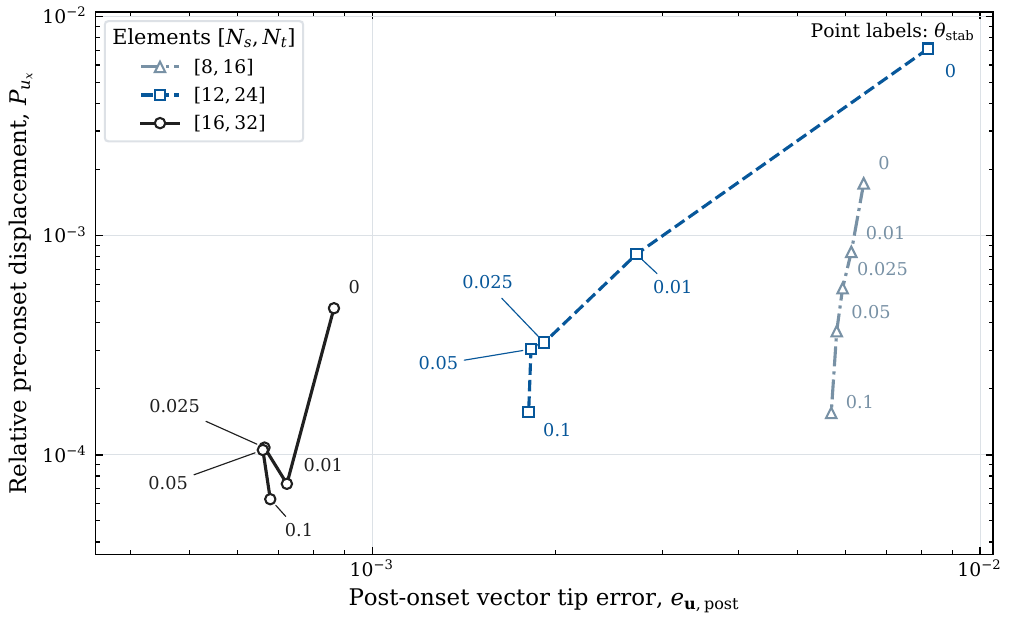}
  \caption{Delayed-load pre-response--accuracy trade-off on three aligned
  meshes. The ordinate is the normalized pre-onset transverse
  displacement; the abscissa is the post-onset relative vector
  tip-displacement error against the finite working reference. Point
  labels denote \(\theta_{\mathrm{stab}}\), connected in increasing order.}
  \label{fig:num_q2_causality_tradeoff}
\end{figure}

Figure~\ref{fig:num_q2_causality_tradeoff} shows the trade-off on all
three grids. On \([16,32]\), nonzero stabilization reduces the
displacement pre-response by factors between approximately 4.3 and 7.4,
with larger reductions in the momentum indicators. At 0 and 0.05,
\((P_{u_x},P_{\|\mi p\|},P_{\|\mi\ell\|},e_{\mi u,\mathrm{post}})\)
are respectively
\((4.660\times10^{-4},6.974\times10^{-3},4.420\times10^{-2},8.640\times10^{-4})\)
and
\((1.051\times10^{-4},5.374\times10^{-4},8.684\times10^{-5},6.600\times10^{-4})\).
The post-onset error is smallest near 0.05; increasing
the parameter to 0.10 slightly improves displacement pre-response at
the cost of a slightly larger post-onset error. The dependence is not
monotone on every grid. Since the working reference itself uses 0.05,
this comparison supports a practical choice, not an independent
parameter optimum or strict domain-of-dependence causality.

Final active residuals lie between \(3.18\times10^{-14}\) and
\(1.29\times10^{-13}\). The changed histories thus coexist with closely
satisfied assembled equations. These coefficient residuals are distinct
from strong compatibility defects and do not establish an additional
momentum-conservation property.

%------------------------------------------------------------
\subsection{Covariance under constant superposed spatial rotations}
\label{subsec:num_objectivity}

Independently solved base and rotated problems use \([N_s,N_t]=[4,8]\),
\(Q_1/Q_1\) and \(Q_2/Q_2\), each with full shear integration or SRI.
The common material data, \(t_{\mathrm{end}}=0.35\), and
\(\theta_{\mathrm{stab}}=0.05\) apply. With
\(\ten Q=\exp(0.9\VPTen{\mi a})\),
\(\mi a=(1,2,-1)\tp/\sqrt6\), only spatial configurations, boundary
data and loads are rotated. The material director basis and section
tensors remain fixed, so the comparison is
\(\ten R_2=\ten Q\ten R_1\).

Let \(\xi=\min\{1,\max\{0,t/t_{\mathrm{end}}\}\}\) and
\(\chi(t)=\xi^2(3-2\xi)\). The base end loads are
\[
 \bar{\mi t}_1(L,t)=10^{-3}\chi(t)(1,0.35,0.15)\tp,
 \qquad
 \bar{\mi m}_1(L,t)=5\times10^{-4}\chi(t)(0.35,-0.25,1)\tp,
\]
with unnormalized direction vectors. The rotated loads are
\(\ten Q\bar{\mi t}_1\) and \(\ten Q\bar{\mi m}_1\); distributed
loads vanish. For \(\ten Q_1=\ten I\), \(\ten Q_2=\ten Q\), the
initial traces and global starting histories are
\((\ten Q_i\mi r_0,\ten Q_i,\mi0,\mi0)\). The left-end placement
and orientation remain fixed and terminal outflow is free.

After undoing \(\ten Q\), the largest nodal placement, rotation,
linear-momentum and angular-momentum discrepancies across the four
pairs are \(7.811\times10^{-16}\), \(5.192\times10^{-15}\,\mathrm{rad}\),
\(9.140\times10^{-16}\), and \(2.301\times10^{-19}\), respectively.
The rotation measure is
\(\|\operatorname{Log}(\ten Q\tp\ten R_2\ten R_1\tp)\|\).
All eight solves converge in three Newton corrections; the largest
active residual is \(3.81\times10^{-12}\). These finite-state covariance
tests concern constant rotations; they do not assess sensitivity to the
Newton update path or prove invariance for arbitrary states or
time-dependent observers.

%------------------------------------------------------------
\subsection{Mesh sensitivity of continuous equal-order simplex spaces}
\label{subsec:num_simplex_stability}

The manufactured problem of Subsection~\ref{subsec:num_manufactured_q2}
is repeated with continuous equal-order \(P_2/P_2\) triangles. Each
space--time rectangle is split along slash, backslash or alternating
diagonals, retaining the global nodes of the corresponding \(Q_2\)
grid. The amplitudes, four initial traces, loads and final time remain
unchanged; the main sequence starts from a rest iterate. On oblique
interior facets, the completion of
Appendix~\ref{subsec:app_non_aligned_facets} replaces one-sided spatial
traces by a common numerical co-normal transfer. Opposite outward
numerical fluxes cancel, although the two physical stress traces need
not coincide. Both triangles inherit their parent slab's scale
\(\tau_e=\theta_{\mathrm{stab}}\Delta t_n\), with
\(\theta_{\mathrm{stab}}=0.05\) and \(\alpha_F=0.5\).

Table~\ref{tab:num_q2_p2_contrast} shows large, pattern-sensitive
relative rotation and angular-momentum errors, despite final active
residuals between \(5.42\times10^{-14}\) and \(2.65\times10^{-10}\)
across the 15 simplex runs. The errors use
\eqref{eq:num_relative_field_errors}; \(e_R\) is a relative error,
not an angle in radians. The positive 28-point diagnostic rule and the
13-point assembly rule agree to within a largest relative difference
of \(4.08\times10^{-7}\) in the field errors.

\begin{table}[tbp]
\centering
\small
\setlength{\tabcolsep}{5pt}
\caption{Relative errors at equal global nodal resolutions, normalized
as in \eqref{eq:num_relative_field_errors}. The investigated
\(P_2/P_2\) construction is strongly pattern-sensitive and does not
improve consistently under refinement.}
\label{tab:num_q2_p2_contrast}
\begin{tabular}{c c c c c}
\toprule
mesh & method & pattern & \(e_R\) & \(e_\ell\)\\
\midrule
\([16,32]\) & \(Q_2/Q_2\) & --          & 2.893e-3 & 2.496e-3\\
\([16,32]\) & \(P_2/P_2\) & slash       & 9.086e-3 & 8.647e-2\\
\([16,32]\) & \(P_2/P_2\) & backslash   & 1.642e-2 & 2.007e-1\\
\([16,32]\) & \(P_2/P_2\) & alternating & 1.101e-2 & 3.646e-2\\
\midrule
\([28,56]\) & \(Q_2/Q_2\) & --          & 5.862e-4 & 7.385e-4\\
\([28,56]\) & \(P_2/P_2\) & slash       & 2.196e0  & 9.847e1\\
\([28,56]\) & \(P_2/P_2\) & backslash   & 1.911e0  & 8.415e1\\
\([28,56]\) & \(P_2/P_2\) & alternating & 5.845e-1 & 2.620e1\\
\midrule
\([32,64]\) & \(Q_2/Q_2\) & --          & 3.978e-4 & 5.545e-4\\
\([32,64]\) & \(P_2/P_2\) & slash       & 1.456e0  & 6.605e1\\
\([32,64]\) & \(P_2/P_2\) & backslash   & 2.252e0  & 1.035e2\\
\([32,64]\) & \(P_2/P_2\) & alternating & 5.299e0  & 2.350e2\\
\bottomrule
\end{tabular}
\end{table}

On \([8,16]\), Q2 and slash-P2 controls compare rest and interpolated-exact
initial iterates: their relative rotation and angular-momentum errors
differ by at most \(4.59\times10^{-6}\) relative to each other. Reducing
both manufactured amplitudes by \(10^3\), with the tolerance reduced
from \(10^{-8}\) to \(10^{-11}\), changes these relative errors by at
most \(4.07\times10^{-5}\); no normalization safeguard is active.
These planar coarse-grid controls do not establish fine-grid
initialization independence or assess path sensitivity for noncommuting
three-dimensional rotations.

The refinement study supports the time-aligned \(Q_2/Q_2\)
construction as the numerical baseline. Developing a reliable
non-aligned discretization for the present formulation requires
further work on the choice of spaces and stabilization.

\FloatBarrier

% Version 8: spring main text with completed M2 controls.
% Short controls at 25 ms; within-method and cross-method comparison at 500 ms.
% Working draft: complete H5/M8 history is pending. Long illustration: five runs to 2 s.
\subsection{A helical spring coupled to a rigid body}
\label{subsec:num_spring_rigid_body}

This example combines a precurved beam with translational and rotational
rigid-body motion. Its geometry and material scales are motivated by the
Wilberforce-pendulum example of Harsch et al.~\cite{harsch_finite_2021}.
The initially undeformed spring, \(2\,\mathrm{mm}\) pitch and prescribed
force ramp define a separate initial-boundary-value problem from their
gravity-driven, predeformed pendulum. Short-interval refinement controls
are followed by a half-second comparison during appreciable motion and
a two-second illustration.

\paragraph{Model and loading.}
The spring has 20 turns, coil radius \(R_c=16\,\mathrm{mm}\), pitch
\(2\,\mathrm{mm}\) per turn and circular wire diameter \(d=1\,\mathrm{mm}\).
The material data are \(E=206\times10^9\,\mathrm{Pa}\),
\(G=81.5\times10^9\,\mathrm{Pa}\) and
\(\rho=7850\,\mathrm{kg\,m^{-3}}\), with
\(A_{\mathrm{cs}}=\pi d^2/4\), \(I_1=I_2=\pi d^4/64\) and polar
second moment \(I_1+I_2\). The body has mass \(m_C=0.469\,\mathrm{kg}\)
and body-frame centroidal inertia
\(\ten J_C=\operatorname{diag}(1.468,1.468,1.247)
\times10^{-4}\,\mathrm{kg\,m^2}\).
The upper spring end is clamped. A massless weld connects the lower end
to the body, whose centre-to-attachment vector is
\(\mi a_C=(0.016,0,0)\tp\,\mathrm{m}\).
With \(h_c=(2\,\mathrm{mm})/(2\pi)\) and
\(c_h=\sqrt{R_c^2+h_c^2}\), the reference centreline is
\[
 \mi r_0(s)=\bigl(R_c(\cos\varphi-1),-R_c\sin\varphi,-h_c\varphi\bigr)\tp,
 \qquad \varphi=s/c_h\in[0,40\pi].
\]
The reference directors are its normal, binormal and tangent, with no
additional twist. The body director matrix is
\(\ten\Lambda_C(t)=\sum_{i=1}^3\mi b_i(t)\otimes\mi e_i\).
Initially, \(\ten\Lambda_C(0)=\ten I\),
\(\mi r_C=\mi r_0(L)-\mi a_C\), and both
components are undeformed and at rest. The first-window Newton iterate
extends this state constantly over time; later windows extend the
accepted terminal configuration, beam momenta and body velocities,
fixing only incoming traces and the clamp.
The centre force uses the previously defined quintic ramp,
\[
 \mi f_C(t)=-4q(t/t_r)\mi e_3\,\mathrm{N},
 \qquad t_r=0.0125\,\mathrm{s},
\]
with no additional couple, gravity or contact.
The beam uses \(\theta_{\mathrm{stab}}=0.05\); the body has no added
time-upwind stabilization. Space--time assembly uses four and five
Gauss points per parametric direction for \(p_t=2\) and 3,
respectively; beam energy at fixed times is evaluated with four spatial
Gauss points per element.

\paragraph{Coupling and midpoint comparator.}
The body uses the displacement--velocity form of the rotationless
space--time formulation \cite{hesch_space-time_2024}. Its centre position,
directors, independent velocities and six metric multipliers use the
beam's temporal mesh and degree. The metric condition
\(\ten\Lambda_C\tp\ten\Lambda_C=\ten I\) and velocity compatibility are imposed
weakly in time. Six further multiplier fields impose the weld conditions
weakly in time,
\[
 \mi r(L,t)=\mi r_C(t)+\ten\Lambda_C(t)\mi a_C,
 \qquad
 \ten\Lambda(L,t)=\ten\Lambda_C(t)\ten\Lambda_0(L).
\]
Body and multiplier coefficients are updated additively; the beam
rotation follows Subsection~\ref{subsec:disc_fields}.

The comparator combines the implicit midpoint rule with an established
director-based constrained spatial formulation
\cite{betsch_frameindifferent_2002,leyendecker_objective_2006}.
Placement and directors are interpolated componentwise, with
orthonormality enforced algebraically at the spatial nodes and a
consistent Cartesian mass matrix. The body is condensed into the
beam's end-node variables through
\(\ten\Lambda_C=\ten\Lambda_{\mathrm{MP}}(L)\ten\Lambda_0(L)\tp\) and
\(\mi r_C=\mi r_{\mathrm{MP}}(L)-\ten\Lambda_C\mi a_C\); its inertia and
force are pulled back through this fixed linear map.
Thus each method retains its native rotation, constraint and inertia
discretization. Their comparison assesses the complete discretizations
of the same mechanical problem at a stated spatial resolution.

\paragraph{Short-interval refinement controls.}
On \(0\le t\le T=0.025\,\mathrm{s}\), the common space--time reference
uses \(N_s^{\mathrm{ref}}=1280\),
\(p_s^{\mathrm{ref}}=p_t^{\mathrm{ref}}=3\),
\(\Delta t_{\mathrm{ref}}=4.8828125\times10^{-5}\,\mathrm{s}\) and
\(H_{\mathrm{ref}}=9.765625\times10^{-5}\,\mathrm{s}\).
We compare body-centre position \(\mi r_C\), linear momentum \(\mi P_C\),
director matrix \(\ten\Lambda_C\) and centroidal spin
\(\mi S_C=\mi H_O-\mi r_C\times\mi P_C\), where \(\mi H_O\) is
angular momentum about the fixed origin:
\begin{equation}
 \begin{aligned}
 e_r(t)&=\|\mi r_C^h(t)-\mi r_C^{\mathrm{ref}}(t)\|_2,
 &e_P(t)&=\|\mi P_C^h(t)-\mi P_C^{\mathrm{ref}}(t)\|_2,\\
 e_\Lambda(t)&=\|\ten\Lambda_C^h(t)-\ten\Lambda_C^{\mathrm{ref}}(t)\|_F,
 &e_S(t)&=\|\mi S_C^h(t)-\mi S_C^{\mathrm{ref}}(t)\|_2.
 \end{aligned}
 \label{eq:num_spring_endpoint_errors}
\end{equation}
Their units are \(\mathrm{m}\), \(\mathrm{kg\,m\,s^{-1}}\), one and
\(\mathrm{kg\,m^2\,s^{-1}}\); \(e_\Lambda\) is a matrix difference.
The short-interval study evaluates these quantities at the shared
endpoint \(T\), without interpolation. Newton tolerances are
\(10^{-10}\) for space--time and \(10^{-7}\) for midpoint.

\begin{figure}[!htbp]
 \centering
 \includegraphics[width=\textwidth]{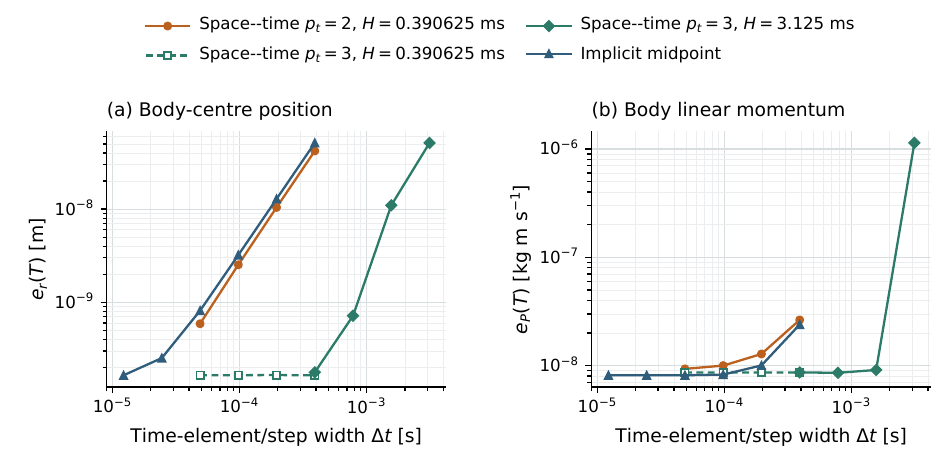}
 \caption{Common-reference endpoint position and linear-momentum errors
at \(T=0.025\,\mathrm{s}\), \(N_s=320\), \(p_s=3\).
The abscissa is the midpoint step size or the space--time element width.
The space--time window length is \(H=0.390625\,\mathrm{ms}\), or
\(3.125\,\mathrm{ms}\) for the additional cubic sequence.
Both spatial and temporal errors contribute.}
 \label{fig:num_spring_time_absolute}
\end{figure}

At \(N_s=320\), \(p_s=3\), the main time sequence uses
\(H=0.390625\,\mathrm{ms}\), \(m=1,2,4,8\) elements per window
and \(\Delta t=H/m\), for \(p_t=2,3\).
Figure~\ref{fig:num_spring_time_absolute} shows approximately
second-order position decay for \(p_t=2\), with successive rates
2.014, 2.041 and 2.104. The cubic sequence lies near its spatial
plateau; the additional sequence with \(H=3.125\,\mathrm{ms}\) and
the same \(m\) values resolves decay towards that plateau, with rates
2.224, 3.936 and 2.015. These nonuniform cubic rates do not establish
an asymptotic temporal order. The midpoint sequence uses
64--2048 steps; successive midpoint position differences have rates
1.998--2.000 as differences to the common reference level off.

The finest reference accepts all 256 windows. The two 1280-element
references differ by a factor two in time-element width and window
length; their endpoint differences are
\(2.01\times10^{-12}\,\mathrm{m}\) in position and
\(5.19\times10^{-11}\,\mathrm{kg\,m\,s^{-1}}\) in linear momentum.
At 2048 midpoint steps, increasing \(N_s\) from 320 to 640 and 1280
reduces \(e_P(T)\) from \(8.131\times10^{-9}\) to
\(2.418\times10^{-10}\) and \(7.600\times10^{-11}\,\mathrm{kg\,m\,s^{-1}}\).
Tightening midpoint tolerances to \(10^{-8}\) at 256 steps and
\(5\times10^{-8}\) at 2048 steps changes position
by at most \(1.47\times10^{-13}\,\mathrm{m}\).
These controls support the spatial interpretation of the plateau;
the finest values remain sensitive to the finite numerical reference.

\begin{figure}[!htbp]
 \centering
 \includegraphics[width=0.90\textwidth]{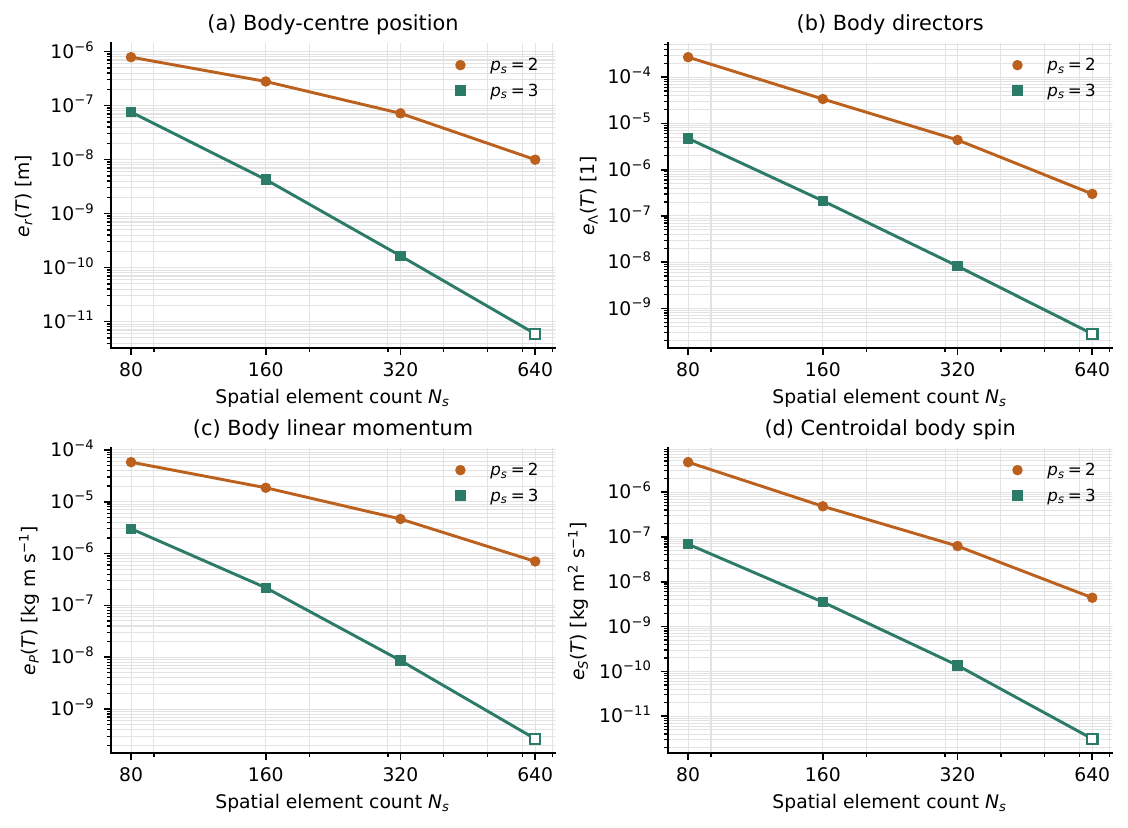}
 \caption{Spatial refinement against the common space--time reference
at \(T=0.025\,\mathrm{s}\), with \(p_t=3\) and
\(\Delta t=9.765625\times10^{-5}\,\mathrm{s}\).
The endpoint norms follow \eqref{eq:num_spring_endpoint_errors}.
Open markers identify the reference-sensitive finest cubic results.}
 \label{fig:num_spring_space_absolute}
\end{figure}

The separate spatial study uses \(p_s=2,3\),
\(N_s=80,160,320,640\), \(p_t=3\),
\(\Delta t=9.765625\times10^{-5}\,\mathrm{s}\) and
\(H=3.90625\times10^{-4}\,\mathrm{s}\).
All four endpoint measures decrease
(Figure~\ref{fig:num_spring_space_absolute}); the finest cubic position
difference is \(5.94\times10^{-12}\,\mathrm{m}\), comparable to the
reference sensitivity. These are finite-sequence endpoint results.

\paragraph{Temporal refinement over half a second.}
To assess resolution during appreciable motion, we extend the comparison
to \(T=0.5\,\mathrm{s}\), keeping \(N_s=320\) and \(p_s=3\) for
both methods. Space--time uses \(p_t=3\), windows
\(H=5,10,15,20\,\mathrm{ms}\) and \(m=4,8\) elements per window,
with nominal width \(h_t=H/m\); the final \(15\,\mathrm{ms}\) window
is shortened to \(5\,\mathrm{ms}\). Its Newton tolerance is \(10^{-8}\).
Midpoint steps are \(0.1,0.5,1,2,5\,\mathrm{ms}\); these runs use
tolerance \(10^{-7}\) up to their accepted restart states between
420 and 430 ms and \(2\times10^{-7}\) thereafter.
Only completed trajectories enter full-interval norms.

\begin{figure}[!t]
 \centering
 \includegraphics[width=0.90\textwidth]{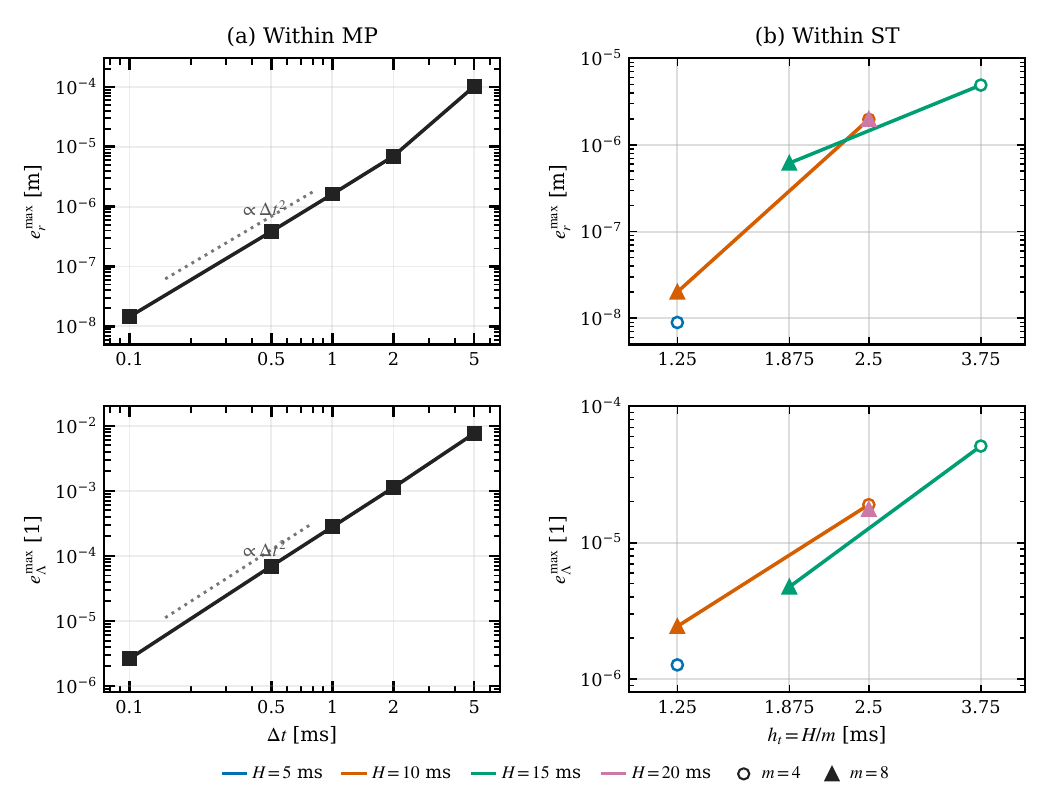}
 \caption{Temporal self-comparisons over \(0\le t\le0.5\,\mathrm{s}\)
at \(N_s=320\), \(p_s=3\): midpoint against the
\(25\,\mu\mathrm{s}\) midpoint solution (left), and space--time
with \(p_t=3\) against the \(H=5\,\mathrm{ms}\), \(m=8\)
space--time solution (right).
Shown are the sampled maxima \eqref{eq:num_spring_history_differences}.
For ST, colours indicate \(H\); circles and triangles indicate
\(m=4\) and 8. Lines connect ST points only at fixed \(H\).
The ST reference has zero self-difference and is omitted from the
logarithmic axes. The \(H=20\,\mathrm{ms}\), \(m=4\) trajectory
ends at 440 ms and is excluded.}
 \label{fig:num_spring_500ms_comparison}
\end{figure}

Each method is compared with its own finest completed trajectory:
the \(25\,\mu\mathrm{s}\) midpoint solution and the space--time solution
with \(H=5\,\mathrm{ms}\), \(m=8\), respectively.
We evaluate the sampled maxima shown in
Figure~\ref{fig:num_spring_500ms_comparison},
\begin{equation}
 e_z^{\max}=\max_{t\in\mathcal G}e_z(t),
 \qquad \mathcal G=\{0,0.01,\ldots,0.5\}\,\mathrm{s},
 \qquad z\in\{r,\Lambda\},
 \label{eq:num_spring_history_differences}
\end{equation}
using the position and director-matrix norms in
\eqref{eq:num_spring_endpoint_errors}.
These 51 times are native to every compared solution.
No state interpolation or rotation projection is used.

\emph{Within midpoint.}
The comparison solution uses tolerance \(2\times10^{-7}\) throughout.
Independent 100, 50 and 25-\(\mu\mathrm{s}\) runs give
successive-difference rates 1.999 for position and 2.008 for directors
on their shared 100-\(\mu\mathrm{s}\) grid.
The 50/25-\(\mu\mathrm{s}\) differences are at most
\(2.98\times10^{-9}\,\mathrm{m}\) in position and \(5.23\times10^{-7}\)
in directors on all shared native times. Under a second-order
leading-error model, the remaining temporal error of the finest run
is estimated as one third of these differences.
Changing its tolerance to \(10^{-6}\) changes position by at most
\(3.03\times10^{-12}\,\mathrm{m}\).

\emph{Within space--time.}
The comparison solution accepts all 100 windows.
At fixed \(H=10\) and \(15\,\mathrm{ms}\), increasing \(m\) from
4 to 8 reduces both sampled differences.
For \(m=8\), reducing \(H\) from 20 through 15 to
\(10\,\mathrm{ms}\) decreases the sampled position maxima from
\(1.98\times10^{-6}\) to \(2.01\times10^{-8}\,\mathrm{m}\)
and the director maxima from \(1.76\times10^{-5}\) to
\(2.44\times10^{-6}\).
The \(H=5\,\mathrm{ms}\), \(m=4\) trajectory differs from the
comparison solution by at most \(8.95\times10^{-9}\,\mathrm{m}\)
in position and \(1.27\times10^{-6}\) in directors on the same grid.

\begin{samepage}
\emph{Difference between the finest solutions.}
At the common endpoint \(T=0.5\,\mathrm{s}\), the finest midpoint
and space--time solutions differ by
\[
 \begin{aligned}
 \|\mi r_C^{\mathrm{ST}}(T)-\mi r_C^{\mathrm{MP}}(T)\|_2
   &\simeq 1.330\times10^{-6}\,\mathrm{m},\\
 \|\ten\Lambda_C^{\mathrm{ST}}(T)
       -\ten\Lambda_C^{\mathrm{MP}}(T)\|_F
   &\simeq 5.228\times10^{-5}.
 \end{aligned}
\]
These are absolute differences of the native body observables.
At the fixed spatial resolution, they retain the methods' distinct
rotation, constraint and inertia discretizations as well as their
remaining temporal errors.
\end{samepage}

\begin{figure}[!htbp]
 \centering
 \includegraphics[width=.90\linewidth]{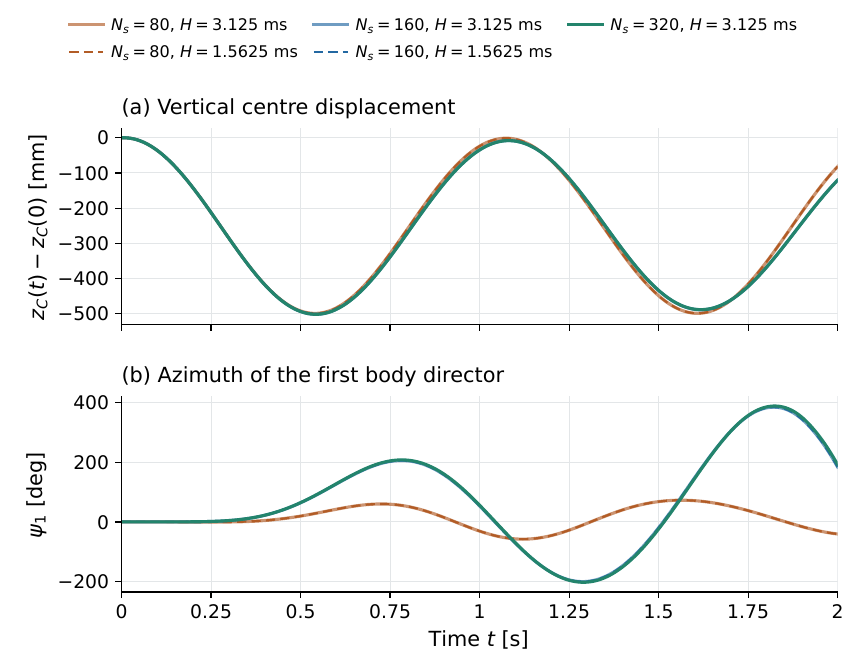}
 \caption{Vertical centre displacement and unwrapped first-director
azimuth over \(0\le t\le2\,\mathrm{s}\) for the five space--time
runs. The time refinements at \(N_s=80\) and 160 nearly overlap.
The director's projection onto the \(xy\) plane is nonsingular at
every recorded sample.}
 \label{fig:spring_motion}
\end{figure}

\paragraph{Motion over two seconds.}
Five space--time runs use \(p_s=p_t=3\), two time elements per window,
\(\Delta t=H/2\) and Newton tolerance \(10^{-8}\):
\(H=3.125\,\mathrm{ms}\) at \(N_s=80,160,320\), and
\(H=1.5625\,\mathrm{ms}\) at \(N_s=80,160\).
All 640 or 1280 windows are accepted.
Figure~\ref{fig:spring_motion} shows vertical centre displacement and
the unwrapped azimuth of the first body director
\(\mi b_1=\ten\Lambda_C\mi e_1\).
At paired native times, halving \(H\) changes position by at most
\(0.1603\) and \(0.5164\,\mu\mathrm{m}\) for \(N_s=80\) and 160;
the corresponding director differences are
\(5.99\times10^{-5}\) and \(1.43\times10^{-4}\).
Spatial refinement from 80 to 160 and from 160 to 320 elements at
\(H=3.125\,\mathrm{ms}\) gives maximum position differences
\(38.53\) and \(0.3280\,\mathrm{mm}\), and director differences
2.8284 and 0.1950. Spatial differences decrease markedly, while the
remaining orientation sensitivity is appreciable.

\begin{figure}[!htbp]
 \centering
 \includegraphics[width=.90\linewidth]{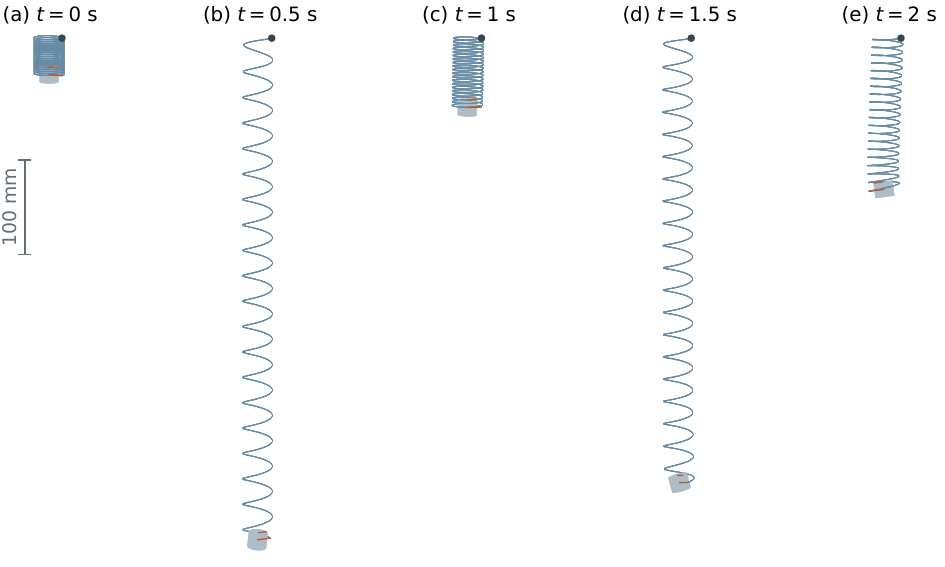}
 \caption{Configurations at \(t=0,\ 0.5,\ 1,\ 1.5,\ 2\,\mathrm{s}\)
for \(N_s=320\), \(H=3.125\times10^{-3}\,\mathrm{s}\).
All orthographic views share a physical scale and are separated horizontally.
The beam uses 961 native spatial nodes without deformation amplification.
The cylinder (radius \(10\,\mathrm{mm}\), height \(16\,\mathrm{mm}\))
is a schematic body contour in the computed pose. Orange lines indicate
the attachment arm and a radial orientation marker.}
 \label{fig:spring_configurations}
\end{figure}

The total energy \(E=E_h+T_C\) combines
\eqref{eq:num_section_energy} with the body's kinetic energy,
\[
 T_C=\tfrac12m_C\|\mi v_C\|^2+
 \tfrac12\operatorname{tr}(\ten V_C\ten E_C\ten V_C\tp),
 \qquad \ten E_C=\tfrac12\operatorname{tr}(\ten J_C)\ten I-\ten J_C.
\]
It uses the independent body velocities \(\mi v_C,\ten V_C\).
Work instead uses the derivative of the computed centre position,
\[
 W_C(t)=\int_0^t\mi f_C(\xi)\cdot\dot{\mi r}_C(\xi)\,\d\xi,
 \qquad d_E(t)=E(t)-E(0)-W_C(t),
\]
integrated from the native temporal polynomials and split at ramp
completion and element boundaries. The sampled \(\max|d_E|\) is
approximately \(1.486\times10^{-2}\), \(4.31\times10^{-4}\) and
\(1.085\times10^{-6}\,\mathrm{J}\) for \(N_s=80,160,320\),
respectively; the last is \(5.409\times10^{-5}\%\) of
\(\max|W_C|\).
Figure~\ref{fig:spring_configurations} shows the \(N_s=320\) motion.

\FloatBarrier

\section{Conclusions}
\label{sec:conclusions}

The formulation describes geometrically exact Cosserat-beam dynamics
on a directed world sheet with prescribed absolute time and a fixed,
possibly curved and twisted reference configuration. Spatial resultants
and temporal momenta enter common force and moment fluxes. Their
co-normal boundary operator treats spatial natural data, temporal
inflow and free terminal outflow within the same framework: at the
fixed final time, the outgoing momenta are identified with the interior
traces without prescribing the terminal configuration.

Independent momenta are introduced in the continuous mixed action.
The numerical method discretizes the continuously rearranged
first-order equations using time-aligned tensor-product spaces,
future-directed Petrov--Galerkin tests and multiplicative rotation
updates. The distinction between continuous rearrangement and discrete
projection is retained through separate momentum--velocity
compatibility and mechanical balance diagnostics.

The reported tests support the aligned construction. Exact rigid motion
checks temporal momentum transfer, and the smooth manufactured solution
shows decreasing primary-field, resultant and compatibility errors along
the reported refinement sequence. The independently solved base and
rotated problems agree at the nodes to near roundoff after undoing the
constant superposed rotation. Spatial shear and single-mode controls
distinguish spatial effects from temporal errors, while an over-resolved implicit-midpoint
solution provides a complementary comparison of sampled tip histories.
Agreement at matching temporal nodal spacings does not imply identical
finite-rotation spatial discretizations or equal computational cost.

The spring--rigid-body example extends these checks to a curved
reference beam. On the short common interval, the endpoint-position
errors show approximately second-order decay for the quadratic
space--time sequence and the initial midpoint refinements.
The additional coarse cubic sequence approaches the fine-grid plateau
with nonuniform rates, without establishing an asymptotic temporal order.
Spatial refinement provides a separate accuracy check. Over half a
second at \(N_s=320\), midpoint self-differences retain approximately
second-order decay, while space--time self-differences decrease
nonuniformly under element and window refinement. At
\(T=0.5\,\mathrm{s}\), the finest solutions differ by
\(1.330\,\mu\mathrm{m}\) in body position and \(5.228\times10^{-5}\)
in the director-matrix Frobenius norm. This comparison retains the
native rotation, constraint and inertia discretizations of each method
at the stated finite spatial resolution.
The two-second controls show much smaller changes under temporal
refinement at \(N_s=80\) and 160 than under spatial refinement;
adding \(N_s=320\) markedly reduces the differences between successive
spatial resolutions.

Further work should address
stable compatible or broken momentum spaces, higher-order temporal
accuracy with sufficiently resolved spatial references, and adaptive resolution with specified
configuration and momentum transfer between windows. These extensions
require the same separation of approximation error, algebraic
convergence, compatibility and mechanical balance.

\section*{Acknowledgments}
Support for this research was provided by the Deutsche Forschungsgemeinschaft
(DFG), Germany under grant HE5943/27-1. This support is gratefully acknowledged.

\section*{Declaration of generative AI and AI-assisted technologies in the manuscript preparation process}

During the preparation of this work, the authors used ChatGPT (OpenAI)
to assist with language editing, mathematical
consistency checks, and analysis of diagnostic
output. The authors are responsible for independently checking all
AI-assisted material and take full responsibility for the derivations,
computational results, interpretation, and conclusions presented in
this article.

\section*{Conflict of interest}

The authors declare that they have no known competing financial
interests or personal relationships that could have appeared to
influence the work reported in this paper.

\section*{Data availability}

The implementation forms part of the authors' in-house finite-element
framework and is not distributed as a standalone public code package.
The data that support the findings of this study are available from the
authors upon reasonable request.

\appendix
% Version 8 (26 October 2026): integrate former A.1 into Section 3.5 and streamline the temporal-completion discussion.
%============================================================
\section{Technical details, alternative spaces, and non-aligned facets}
\label{app:stabilization_details}
\label{app:compatible_momentum_spaces}

This appendix collects the algebraic steps that are useful for verification and implementation but would obscure the main presentation.
The main method of Section~\ref{sec:spacetime_discretization} uses aligned continuous equal-order tensor-product spaces, with \(Q_2/Q_2\) as the quadratic baseline; other degrees are specified with each experiment.
We derive the temporal boundary completion and the element residuals of the aligned method.
The balance diagnostics are followed by the conservative non-aligned-facet extension used in the simplex diagnostics.
Under the stated trace and stabilization assumptions, that extension is consistent and uses pairwise conservative numerical fluxes. These properties do not imply elementwise balance for a continuous test space or mesh-uniform stability of continuous equal-order simplex spaces.
The appendix concludes with alternative momentum spaces.

%------------------------------------------------------------
\subsection{Temporal integration by parts and flux completion}
\label{subsec:app_temporal_completion}

Integration by parts of \(\mi p\cdot\mi\eta_{,t}\) and \(\mi\ell\cdot\mi\theta_{,t}\), followed by multiplication of the balance rows by \(-1\), places the temporal derivatives on \(\mi p\) and \(\mi\ell\). The resulting equations remain tested and integrated over space and time.
The endpoint terms are combined with the temporal co-normal external work from \eqref{eq:worldsheet_temporal_natural_work}.
The temporal contribution of the Galerkin core is
\begin{equation}
  \mathcal G_{h,t}^{0}
  :=\mathcal G_{h,t}^{0,\mathrm{ext}}
    +\mathcal G_{h,t}^{0,\mathrm{int}}.
  \label{eq:disc_temporal_functional_split}
\end{equation}
On the external temporal faces,
\begin{align}
  \mathcal G_{h,t}^{0,\mathrm{ext}}
  :=&\int_{\Sigma_T}
  \Big[
    (\bar{\mi p}^{+}-\mi p_h)\cdot\mi\eta_h
    +(\bar{\mi\ell}^{+}-\mi\ell_h)\cdot\mi\theta_h
  \Big]\d s
  \nonumber\\
  &+\int_{\Sigma_0}
  \Big[
    (\mi p_h-\bar{\mi p}^{-})\cdot\mi\eta_h
    +(\mi\ell_h-\bar{\mi\ell}^{-})\cdot\mi\theta_h
  \Big]\d s.
  \label{eq:disc_external_temporal_residual}
\end{align}
With the strongly imposed incoming traces and the outflow choice \eqref{eq:disc_terminal_outflow_flux}, both mismatch terms vanish by matching the momentum traces.
After the Petrov--Galerkin perturbation, vanishing unperturbed tests alone would not justify discarding the temporal work; the corresponding cancellation is derived below.

For a formulation with broken momenta across a constant-time interface,
\begin{align}
  \mathcal G_{h,t}^{0,\mathrm{int}}
  :=\sum_{F\in\mathcal F_{h,t}^{\mathrm{int}}}\int_F
  \Big[
    (\mi p_h^+-\widehat{\mi p}_h^- )\cdot\mi\eta_h^+
    +(\mi\ell_h^+-\widehat{\mi\ell}_h^- )\cdot\mi\theta_h^+
  \Big]\d s.
  \label{eq:disc_temporal_jump_terms}
\end{align}
Here \(+\) denotes the receiving slab and the one-sided choice is the predecessor's own outgoing trace,
\(\widehat{\mi p}_h^-=\mi p_h^-\) and \(\widehat{\mi\ell}_h^-=\mi\ell_h^-\).
The donor-side mismatch therefore vanishes. A modified or projected transfer would require its own two-sided flux treatment.
For the globally continuous momenta used in the main method, the incoming trace equals the preceding outgoing trace, so \(\mathcal G_{h,t}^{0,\mathrm{int}}=0\).
In a windowed solve this is a prescribed transfer between consecutive solution windows, as described in Subsection~\ref{subsec:disc_newton}; it does not identify that solve with a single monolithic discretization.

For the perturbed tests, the external temporal work is
\begin{align}
  \delta\mathcal A_{t,h}^{\mathrm{ext},\star}
  =&-\int_{\Sigma_T}
    \left(
      \bar{\mi p}^{+}\cdot\mi\eta_h^{\star}
      +\bar{\mi\ell}^{+}\cdot\mi\theta_h^{\star}
    \right)\d s
  \nonumber\\
  &+\int_{\Sigma_0}
    \left(
      \bar{\mi p}^{-}\cdot\mi\eta_h^{\star}
      +\bar{\mi\ell}^{-}\cdot\mi\theta_h^{\star}
    \right)\d s.
  \label{eq:disc_stabilized_temporal_external_action}
\end{align}
The cancellation with the endpoint terms generated by temporal integration by parts is
\begin{equation}
  \int_{t_n}^{t_{n+1}}
    \mi p\cdot\partial_t\mi\eta_h^{\star}\,\d t
  -\left[\mi p\cdot\mi\eta_h^{\star}\right]_{t_n}^{t_{n+1}}
  =-\int_{t_n}^{t_{n+1}}
    \mi p_{,t}\cdot\mi\eta_h^{\star}\,\d t,
  \label{eq:disc_stabilized_temporal_cancellation}
\end{equation}
and analogously for \(\mi\ell\) and \(\mi\theta_h^{\star}\).
This identity is applied on each time slab with its own one-sided starred traces.
For continuous momenta, matching the numerical and physical traces makes each local flux mismatch vanish separately, as detailed in Subsection~\ref{subsec:disc_stabilization}.
Thus the free upper face is an outflow boundary; it is not a deleted boundary equation.

%------------------------------------------------------------
\subsection{Aligned Petrov--Galerkin completion and element residuals}
\label{subsec:app_complete_pg_decomposition}
\label{subsec:app_element_residuals}

For the element residuals and balance identities, define on each element
\begin{align}
  \mathcal E_r^h
  &:=\mi p^h_{,t}-\mi n^h_{,s}-\mi f^h,
  &
  \mathcal E_\theta^h
  &:=\mi\ell^h_{,t}-\mi m^h_{,s}
    -\mi r^h_{,s}\times\mi n^h-\mi c^h,
  \label{eq:disc_balance_residuals}\\
  \mathcal E_p^h
  &:=\mi r^h_{,t}-A^{-1}\mi p^h,
  &
  \mathcal E_\ell^h
  &:=\mi\omega^h-(\ten J^{\mathrm{sp},h})^{-1}\mi\ell^h.
  \label{eq:disc_kinematic_residuals}
\end{align}
The spatial derivatives of \(\mi n^h\) and \(\mi m^h\) include variations of the prescribed reference tangent, orientation and section data.

On an element \(\mathcal B_e=\mathcal B_{an}\), set
\begin{equation}
  N_A^{\star,e}:=N_A+\tau_nN_{A,t},
  \qquad
  M_B^{\star,e}:=M_B+\tau_nM_{B,t}.
  \label{eq:app_starred_basis}
\end{equation}
Choosing successively
\(\mi\eta^h=N_A\mi e_i\),
\(\mi\theta^h=N_A\mi e_i\),
\(\mi\pi^h=M_B\mi e_i\), and
\(\mi\rho^h=M_B\mi e_i\) in the direct starred weak form gives
\begin{align}
  \mi R_{r,A}^e
  =&\int_{\mathcal B_e}
  \Big[
    N_A^{\star,e}\mi p^h_{,t}
    +(N_A^{\star,e})_{,s}\mi n^h
    -N_A^{\star,e}\mi f^h
  \Big]\d s\,\d t
  +\mi R_{r,A}^{e,\mathrm{bd}},
  \label{eq:disc_vector_residual_r}\\
  \mi R_{\theta,A}^e
  =&\int_{\mathcal B_e}
  \Big[
    N_A^{\star,e}\mi\ell^h_{,t}
    +(N_A^{\star,e})_{,s}\mi m^h
    -N_A^{\star,e}(\mi r^h_{,s}\times\mi n^h+\mi c^h)
  \Big]\d s\,\d t
  +\mi R_{\theta,A}^{e,\mathrm{bd}},
  \label{eq:disc_vector_residual_theta}\\
  \mi R_{p,B}^e
  =&\int_{\mathcal B_e}
    M_B^{\star,e}\mathcal E_p^h\,\d s\,\d t,
  \label{eq:disc_vector_residual_p}\\
  \mi R_{\ell,B}^e
  =&\int_{\mathcal B_e}
    M_B^{\star,e}\mathcal E_\ell^h\,\d s\,\d t.
  \label{eq:disc_vector_residual_ell}
\end{align}
These are exact-integral expressions. Their numerical assembly uses the quadrature rules specified with the numerical examples.
They are assembled before essential elimination; the solved rows correspond to the free tests specified in Subsection~\ref{subsec:disc_fields}. Constrained rows are not additional equations for the unknown coefficients.
The boundary vectors split as
\begin{equation}
  \mi R_{\bullet,A}^{e,\mathrm{bd}}
  =\mi R_{\bullet,A}^{e,N}
   +\mi R_{\bullet,A}^{e,t,\mathrm{ext}}
   +\mi R_{\bullet,A}^{e,t,\mathrm{int}},
  \qquad \bullet\in\{r,\theta\}.
  \label{eq:disc_boundary_residual_split}
\end{equation}
On an aligned spatial natural face, the prescribed end loads are integrated over material time:
\begin{align}
  \mi R_{r,A}^{e,N}
  =&-\int_{\partial\mathcal B_e\cap\Sigma_r^N}
    N_A^{\star,e}\bar{\mi t}^h\,\d t,
  \label{eq:disc_element_neumann_residual_r}\\
  \mi R_{\theta,A}^{e,N}
  =&-\int_{\partial\mathcal B_e\cap\Sigma_R^N}
    N_A^{\star,e}\bar{\mi m}^h\,\d t.
  \label{eq:disc_element_neumann_residual_theta}
\end{align}
Here \(\bar{\mi t}^h\) and \(\bar{\mi m}^h\) already contain the outward orientation.
These formulas are the material pullbacks of the geometric boundary work in \eqref{eq:worldsheet_material_boundary_fluxes}; no additional metric factor or normal sign is introduced.
For an element adjacent to an external temporal face,
\begin{align}
  \mi R_{r,A}^{e,t,\mathrm{ext}}
  =&\int_{\partial\mathcal B_e\cap\Sigma_T}
    N_A^{\star,e}(\bar{\mi p}^{+}-\mi p_h)\,\d s
  +\int_{\partial\mathcal B_e\cap\Sigma_0}
    N_A^{\star,e}(\mi p_h-\bar{\mi p}^{-})\,\d s,
  \label{eq:disc_element_external_time_residual_r}\\
  \mi R_{\theta,A}^{e,t,\mathrm{ext}}
  =&\int_{\partial\mathcal B_e\cap\Sigma_T}
    N_A^{\star,e}(\bar{\mi\ell}^{+}-\mi\ell_h)\,\d s
  +\int_{\partial\mathcal B_e\cap\Sigma_0}
    N_A^{\star,e}(\mi\ell_h-\bar{\mi\ell}^{-})\,\d s.
  \label{eq:disc_element_external_time_residual_theta}
\end{align}
For a broken-momentum interface,
\begin{align}
  \mi R_{r,A}^{e,t,\mathrm{int}}
  :=&\sum_{\substack{F\in\mathcal F_{h,t}^{\mathrm{int}}\\e=e^+(F)}}
  \int_F
    N_A^{\star,+}(\mi p_h^+-\widehat{\mi p}_h^-)\,\d s,
  \label{eq:disc_element_internal_time_residual_r}\\
  \mi R_{\theta,A}^{e,t,\mathrm{int}}
  :=&\sum_{\substack{F\in\mathcal F_{h,t}^{\mathrm{int}}\\e=e^+(F)}}
  \int_F
    N_A^{\star,+}(\mi\ell_h^+-\widehat{\mi\ell}_h^-)\,\d s.
  \label{eq:disc_element_internal_time_residual_theta}
\end{align}
These terms vanish for continuous momenta with matching numerical traces, independently of the selected equal polynomial degree.

Throughout the following element and facet identities, \(\nu^e=(\nu_s^e,\nu_t^e)\) is the outward normal in the fixed material parameter chart and \(\d\widehat\ell\) is its line measure.
The geometric world-sheet measure \(\d\ell\) is used only before pullback, as in \eqref{eq:worldsheet_conormal_pullback}.
The weak perturbation admits a useful residual interpretation.
For the translational row on one element,
\begin{align}
  &\int_{\mathcal B_e}\tau_n
  \left(
    \mi p^h_{,t}\cdot\mi\eta^h_{,t}
    +\mi n^h\cdot\mi\eta^h_{,st}
    -\mi f^h\cdot\mi\eta^h_{,t}
  \right)\d s\,\d t
  \nonumber\\
  &\qquad=
  \int_{\mathcal B_e}\tau_n\mathcal E_r^h\cdot\mi\eta^h_{,t}\,\d s\,\d t
  +\int_{\partial\mathcal B_e}\tau_n
    (\mi n^h\nu_s^e)\cdot\mi\eta^h_{,t}\,\d\widehat\ell.
  \label{eq:app_weak_strong_identity_r}
\end{align}
The corresponding angular identity is
\begin{align}
  &\int_{\mathcal B_e}\tau_n
  \left(
    \mi\ell^h_{,t}\cdot\mi\theta^h_{,t}
    +\mi m^h\cdot\mi\theta^h_{,st}
    -(\mi r^h_{,s}\times\mi n^h+\mi c^h)\cdot\mi\theta^h_{,t}
  \right)\d s\,\d t
  \nonumber\\
  &\qquad=
  \int_{\mathcal B_e}\tau_n\mathcal E_\theta^h\cdot\mi\theta^h_{,t}\,\d s\,\d t
  +\int_{\partial\mathcal B_e}\tau_n
    (\mi m^h\nu_s^e)\cdot\mi\theta^h_{,t}\,\d\widehat\ell.
  \label{eq:app_weak_strong_identity_theta}
\end{align}
On aligned conforming meshes the direct starred weak form is the primary definition; identities \eqref{eq:app_weak_strong_identity_r}--\eqref{eq:app_weak_strong_identity_theta} are used only for interpretation and for constructing a conservative non-aligned completion.

%------------------------------------------------------------
\subsection{Balance diagnostics and admissible symmetry tests}
\label{subsec:app_balance_defects}

Let \(t_1<t_2\) delimit a time interval of the computed history. The identities below define balance remainders of that history; they are not additional equations imposed by the discrete method.
For linear momentum, set
\[
  \mathcal P^h(t):=\int_0^L\mi p^h(s,t)\,\d s.
\]
The linear-momentum remainder \(\mathcal D_{\mathcal P}^h\) is defined by
\begin{equation}
  \mathcal P^h(t_2)-\mathcal P^h(t_1)
  =\int_{t_1}^{t_2}[\mi n^h]_{s=0}^{s=L}\,\d t
   +\int_{t_1}^{t_2}\int_0^L\mi f^h\,\d s\,\d t
   +\mathcal D_{\mathcal P}^h.
  \label{eq:disc_linear_momentum_balance}
\end{equation}
A constant translation has zero temporal derivative and is unchanged by the PG map, but it yields a discrete conservation argument only when it is an admissible test on the domain under consideration.
On the solution windows considered here, the homogeneous initial trace already excludes every nonzero constant translation and constant spin as a global test, independently of spatial supports.
Localization to a subinterval also requires its own admissible-test argument. Continuity of the momenta alone therefore does not make \(\mathcal D_{\mathcal P}^h\) vanish on every subinterval.
All three balance remainders below use one-sided interior resultant traces at the beam ends.
Replacing them by prescribed Neumann data or weak support reactions changes the diagnostic by the corresponding boundary mismatch and must be stated explicitly.
The total spatial angular momentum is
\begin{equation}
  \mathcal J^h(t)
  :=\int_0^L(\mi r^h\times\mi p^h+\mi\ell^h)(s,t)\,\d s.
  \label{eq:disc_angular_momentum}
\end{equation}
The geometric phase-space variation associated with a constant rigid rotation is
\begin{equation}
  \mathbf z_{\mi\mu}^h
  =\bigl(
      \mi\mu\times\mi r^h,
      \mi\mu,
      \mi\mu\times\mi p^h,
      \mi\mu\times\mi\ell^h
    \bigr),
  \qquad \mi\mu\in\R^3\ \text{constant}.
  \label{eq:disc_rotation_symmetry_generator}
\end{equation}
This describes the action of a rigid rotation on the fields, not automatically an admissible test tuple for the rearranged discrete residual. Essential trace restrictions and the sign changes of the balance rows must be taken into account separately.
The angular-momentum remainder \(\mathcal D_{\mathcal J}^h\) is defined by
\begin{align}
  \mathcal J^h(t_2)-\mathcal J^h(t_1)
  =&\int_{t_1}^{t_2}
    [\mi r^h\times\mi n^h+\mi m^h]_{s=0}^{s=L}\,\d t
  \nonumber\\
  &+\int_{t_1}^{t_2}\int_0^L
    (\mi r^h\times\mi f^h+\mi c^h)\,\d s\,\d t
  +\mathcal D_{\mathcal J}^h,
  \label{eq:disc_angular_momentum_balance_defect}
\end{align}
Thus \(\mathcal D_{\mathcal J}^h\) is the total remainder for the displayed boundary and load terms, not solely the value of the stabilization on a symmetry generator.
Its vanishing, including for \(\theta_{\mathrm{stab}}=0\), requires a separate admissible-test argument and is not asserted here. A beam with support reactions is not an isolated angular-momentum system.
The strong residuals of \eqref{eq:disc_balance_residuals}--\eqref{eq:disc_kinematic_residuals} give the elementwise identity
\begin{align}
  &\partial_t(\mi r^h\times\mi p^h+\mi\ell^h)
   -\partial_s(\mi r^h\times\mi n^h+\mi m^h)
   -\mi r^h\times\mi f^h-\mi c^h
   \nonumber\\
  &\qquad=
   \mi r^h\times\mathcal E_r^h+\mathcal E_\theta^h
   +\mathcal E_p^h\times\mi p^h.
  \label{eq:app_local_angular_remainder}
\end{align}
Thus even the local angular remainder can include the translational momentum--velocity defect.
Summing this identity over elements also requires the one-sided interior flux contributions; the global remainder in \eqref{eq:disc_angular_momentum_balance_defect} does not assume those physical fluxes to be continuous.

The discrete Hamiltonian density and energy are
\begin{align}
  \mathcal H^h
  &:=\frac12A^{-1}\mi p^h\cdot\mi p^h
    +\frac12\mi\ell^h\cdot(\ten J^{\mathrm{sp},h})^{-1}\mi\ell^h
    +\Psi(s,\mi\Gamma^h,\mathcal K^h),
  \label{eq:disc_hamiltonian_density}\\
  \mathcal E^h(t)
  &:=\int_0^L\mathcal H^h(s,t)\,\d s.
  \label{eq:disc_total_energy}
\end{align}
The strains and resultants use the fixed reference representation of \eqref{eq:disc_reference_tangent}--\eqref{eq:disc_strains}; the explicit \(s\)-dependence of \(\Psi\) includes the reference and section data.
Because these data are time independent, the discrete natural reference introduces no additional temporal reference-work term.
Using the momentum-derived rates in the following boundary and load powers, define the energy remainder by
\begin{align}
  \mathcal E^h(t_2)-\mathcal E^h(t_1)
  =&\int_{t_1}^{t_2}
  \left[
    \mi n^h\cdot A^{-1}\mi p^h
    +\mi m^h\cdot(\ten J^{\mathrm{sp},h})^{-1}\mi\ell^h
  \right]_{s=0}^{s=L}\d t
  \nonumber\\
  &+\int_{t_1}^{t_2}\int_0^L
  \left[
    \mi f^h\cdot A^{-1}\mi p^h
    +\mi c^h\cdot(\ten J^{\mathrm{sp},h})^{-1}\mi\ell^h
  \right]\d s\,\d t
  +\mathcal D_{\mathcal E}^h.
  \label{eq:disc_energy_balance_defect}
\end{align}
The momentum-derived rates need not equal \(\mi r^h_{,t}\) and \(\mi\omega^h\) pointwise on a finite mesh. Consequently, this remainder includes any corresponding power mismatch as well as contributions from discretization, trace evaluation, quadrature and the PG perturbation.
In particular, at a stationary support the configuration power vanishes, whereas the momentum-derived power used here need not vanish on a finite mesh.

To make the mixed contributions explicit, set
\[
  \mi v_p^h:=A^{-1}\mi p^h,
  \qquad
  \mi\omega_\ell^h:=(\ten J^{\mathrm{sp},h})^{-1}\mi\ell^h.
\]
On a smooth element the rotational kinetic term satisfies
\begin{equation}
  \partial_t\!\left[\frac12\mi\ell^h\cdot
    (\ten J^{\mathrm{sp},h})^{-1}\mi\ell^h\right]
  =\mi\omega_\ell^h\cdot\mi\ell^h_{,t}
   +\mi\omega^h\cdot(\mi\omega_\ell^h\times\mi\ell^h).
  \label{eq:app_rotational_kinetic_rate}
\end{equation}
The second term accounts for the time-dependent spatial inertia and cannot be dropped for independent fields.
Together with the elastic energy rate, it gives
\begin{align}
  &\mathcal H^h_{,t}
   -\partial_s(\mi n^h\cdot\mi v_p^h
              +\mi m^h\cdot\mi\omega_\ell^h)
   -\mi f^h\cdot\mi v_p^h-\mi c^h\cdot\mi\omega_\ell^h
   \nonumber\\
  &\qquad=
   \mi v_p^h\cdot\mathcal E_r^h
   +\mi\omega_\ell^h\cdot\mathcal E_\theta^h
   +\mi n^h\cdot\partial_s\mathcal E_p^h
   +\mi m^h\cdot\partial_s\mathcal E_\ell^h
   \nonumber\\
  &\qquad\quad
   +\mathcal E_\ell^h\cdot
     (\mi\omega_\ell^h\times\mi\ell^h
       -\mi r^h_{,s}\times\mi n^h).
  \label{eq:app_local_energy_remainder}
\end{align}
This is an elementwise differential identity; a global residual representation must also retain internal flux contributions.
The defining balance \eqref{eq:disc_energy_balance_defect} remains valid without assuming their cancellation.
The energy is evaluated by spatial integration at fixed time, including cuts through element interiors. No constant-energy condition is imposed on those cuts; convergence of the diagnostic is assessed by refinement rather than assumed as a discrete identity.
For accumulated rotation updates, the evaluation points of these cuts must be registered consistently with the rotation field and derivative updates in Subsection~\ref{subsec:disc_fields}.

%------------------------------------------------------------
\subsection{Non-aligned facets: conservative completion and its limitation}
\label{subsec:app_non_aligned_facets}
\label{subsec:app_flux_split}

This subsection records the extension used in the simplex diagnostics, separately from the aligned tensor-product method.
It uses \(\beta=\partial_t\) and simplex subdivisions of the same time slabs, with \(\tau_e=\tau_n\) on every element in slab \(n\).
Thus uncorrected spatial interfaces retain common traces of \(\tau_n\partial_t\mathbf z_h\), and \(\beta\) remains tangential to the spatial Dirichlet boundary.
An arbitrary element-dependent \(\tau_e\), a different direction or more general exterior boundaries would require their own trace and boundary completion.
Let \(F\) be a non-aligned interior facet shared by elements \(+\) and \(-\), with outward material normals \(\nu^+=-\nu^-\).
The outward physical force and moment flux densities with respect to \(\d\widehat\ell\) are
\begin{equation}
  \mathfrak t_{r}^{e}=\mi n^e\nu_s^e-\mi p^e\nu_t^e,
  \qquad
  \mathfrak t_{\theta}^{e}=\mi m^e\nu_s^e-\mi\ell^e\nu_t^e.
  \label{eq:disc_physical_conormal_flux}
\end{equation}
For either flux, define
\begin{equation}
  \widehat{\mathfrak t}_{F}^{+}(\alpha_F)
  =\frac12(\mathfrak t^+-\mathfrak t^-)
   +\frac{\alpha_F}{2}\sigma_F(\mathfrak t^++\mathfrak t^-),
  \qquad
  \widehat{\mathfrak t}_{F}^{-}=-\widehat{\mathfrak t}_{F}^{+},
  \label{eq:app_flux_family}
\end{equation}
where
\[
  \sigma_F:=\operatorname{sign}(\beta\cdot\nu^+),
  \qquad 0\le\alpha_F\le1.
\]
The direct starred weak form already contains the one-sided spatial traces \(\mi n^e\nu_s^e\) and \(\mi m^e\nu_s^e\) from \eqref{eq:app_weak_strong_identity_r}--\eqref{eq:app_weak_strong_identity_theta}.
To replace them by a common numerical co-normal transfer, add on every adjacent element
\begin{align}
  \mathcal C_{F}^{e}
  :=-\int_F\tau_e
  \Big[
    D_\beta\mi\eta_h^e\cdot
      (\mi p^e\nu_t^e+\widehat{\mathfrak t}_{r,F}^e)
    +D_\beta\mi\theta_h^e\cdot
      (\mi\ell^e\nu_t^e+\widehat{\mathfrak t}_{\theta,F}^e)
  \Big]\d\widehat\ell.
  \label{eq:disc_stabilization_oblique_faces}
\end{align}
For a single facet \(F\subset\partial\mathcal B_e\) and elementwise
constant \(\tau_e\), spatial integration by parts gives
\begin{align}
  \mathcal S_{r,e}^{\mathrm{vol,w}}
  +\mathcal C_{r,F}^{e}
  ={}&
  \int_{\mathcal B_e}
    \tau_e\mathcal E_r^h\cdot D_\beta\mi\eta_h^e
    \,\d s\,\d t
  \nonumber\\
  &+
  \int_F
    \tau_e
    \bigl(
      \mathfrak t_{r,h}^{e}
      -\widehat{\mathfrak t}_{r,F}^{e}
    \bigr)
    \cdot D_\beta\mi\eta_h^e
    \,\d\widehat\ell
  \nonumber\\
  &+
  \int_{\partial\mathcal B_e\setminus F}
    \tau_e
    (\mi n^h\nu_s^e)
    \cdot D_\beta\mi\eta_h^e
    \,\d\widehat\ell.
  \label{eq:app_non_aligned_equivalence_r}
\end{align}
Here \(\mathcal C_{r,F}^{e}\) is the translational part of
\(\mathcal C_F^e\), and all physical traces are taken from element \(e\).
The same argument applies to the moment row.
Thus the correction replaces the spatial boundary contribution on
\(F\) by a numerical co-normal flux residual, while retaining the
spatial boundary contributions on
\(\partial\mathcal B_e\setminus F\).
The associated element vectors are
\begin{align}
  \mi R_{r,A}^{e,F}
  =&-\sum_{\substack{F\in\mathcal F_{h,\mathrm{ob}}^{\mathrm{int}}\\F\subset\partial\mathcal B_e}}
  \int_F\tau_eD_\beta N_A^e
    (\mi p^e\nu_t^e+\widehat{\mathfrak t}_{r,F}^{e})\,\d\widehat\ell,
  \label{eq:disc_element_oblique_residual_r}\\
  \mi R_{\theta,A}^{e,F}
  =&-\sum_{\substack{F\in\mathcal F_{h,\mathrm{ob}}^{\mathrm{int}}\\F\subset\partial\mathcal B_e}}
  \int_F\tau_eD_\beta N_A^e
    (\mi\ell^e\nu_t^e+\widehat{\mathfrak t}_{\theta,F}^{e})\,\d\widehat\ell.
  \label{eq:disc_element_oblique_residual_theta}
\end{align}

The flux family satisfies:
\begin{enumerate}
\item \emph{Numerical-flux conservation:}
\(\widehat{\mathfrak t}_{F}^{+}+\widehat{\mathfrak t}_{F}^{-}=0\).
\item \emph{Consistency:}
if \(\mathfrak t^-=-\mathfrak t^+\), then \(\widehat{\mathfrak t}_{F}^{+}=\mathfrak t^+\) for every \(\alpha_F\).
\item \emph{Central limit:}
\(\alpha_F=0\) gives \(\widehat{\mathfrak t}_{F}^{+}=\tfrac12(\mathfrak t^+-\mathfrak t^-)\).
\item \emph{Temporally one-sided limit:}
for \(\alpha_F=1\) and \(\beta\cdot\nu^+\ne0\), the flux is taken from the temporally upstream element. When \(\beta\cdot\nu^+=0\), \(\sigma_F=0\) gives the central flux.
\item \emph{Side-swap invariance:}
interchanging \(+\) and \(-\) reverses the outward sign but not the physical transfer.
\end{enumerate}
The first property is pairwise conservation of the numerical transfer.
It does not by itself impose an independent balance on each element: an elementwise constant indicator is generally not an admissible continuous test.
For exact continuous reference and load data, exact integration and a smooth exact solution, the physical and numerical co-normal traces coincide.
The facet residuals then vanish; the remaining contributions cancel using the stated slabwise trace assumptions and boundary data.
This is the consistency statement for the completed operator, separate from reference approximation, quadrature and stability.

For the force flux, its physical jump decomposes as
\begin{align}
  \mathfrak j_r
  &:=\mathfrak t_r^++\mathfrak t_r^-
    =\mathfrak j_r^{s}+\mathfrak j_r^{t},
  \label{eq:app_force_jump_split}\\
  \mathfrak j_r^{s}
  &:=\nu_s^+(\mi n^+-\mi n^-),
  &
  \mathfrak j_r^{t}
  &:=-\nu_t^+(\mi p^+-\mi p^-).
  \label{eq:app_force_jump_components}
\end{align}
Similarly,
\begin{align}
  \mathfrak j_\theta
  &:=\mathfrak t_\theta^++\mathfrak t_\theta^-
    =\mathfrak j_\theta^{s}+\mathfrak j_\theta^{t},
  \label{eq:app_moment_jump_split}\\
  \mathfrak j_\theta^{s}
  &:=\nu_s^+(\mi m^+-\mi m^-),
  &
  \mathfrak j_\theta^{t}
  &:=-\nu_t^+(\mi\ell^+-\mi\ell^-).
  \label{eq:app_moment_jump_components}
\end{align}
For globally continuous momenta,
\begin{equation}
  \mathfrak j_r^{t}=\mi0,
  \qquad
  \mathfrak j_\theta^{t}=\mi0,
  \label{eq:app_continuous_momentum_jump_vanishes}
\end{equation}
up to roundoff.
The remaining physical jump is generated by the elementwise spatial resultants.
This does not contradict conservation of the numerical flux.

The split identity
\begin{equation}
  \|\mathfrak j\|^2
  =\|\mathfrak j^s\|^2+\|\mathfrak j^t\|^2
   +2\mathfrak j^s\cdot\mathfrak j^t
  \label{eq:app_jump_norm_identity}
\end{equation}
and the mesh-scaled seminorm
\begin{equation}
  \eta_{\mathfrak j}^2
  :=\sum_{F\in\mathcal F_h^{\mathrm{int}}}
    h_F\int_F\|\mathfrak j\|^2\,\d\widehat\ell
  \label{eq:app_scaled_jump_norm}
\end{equation}
are used only as diagnostics.
For the straight facets of the implemented simplex meshes, the length factor is
\[
  h_F:=\sqrt{(\Delta s_F)^2+c_\star^2(\Delta t_F)^2},
  \qquad c_\star>0,
\]
where \((\Delta s_F,\Delta t_F)\) is the endpoint difference and \(c_\star\) is the recorded characteristic speed.
The normals and integration measure in \eqref{eq:app_scaled_jump_norm} remain those of the fixed material chart \((s,t)\).
The choice of chart, \(c_\star\), selected facets and diagnostic quadrature is held fixed as a convention when comparing refinements.
This diagnostic length scale is distinct from the embedding scale \(c_\tau\).

The implemented facet quantities are material co-normal flux densities
with respect to the line measure \(\d\hat\ell\) in the parameter domain
\(\mathcal B_0\).
Let \(F\subset\partial\mathcal B_e\) be an element edge, let
\(\nu=(\nu_s,\nu_t)\) denote its outward unit normal in the parameter
plane, and set \(\Xi_F:=\Xi|_F\).
On the corresponding world-sheet edge \(\Xi(F)\), let
\(\lambda_\mu\) denote the components of the outward unit co-normal
and let \(\d\ell\) be the induced line measure.
As in \eqref{eq:worldsheet_conormal_pullback}, the co-normal density transforms componentwise as
\[
  \Xi_F^*(\lambda_\mu\,\d\ell)
  =
  \sqrt g\,\nu_\mu\,\d\hat\ell,
  \qquad \mu\in\{s,t\}.
\]
Using the flux definitions
\(\sqrt g\,\mathfrak n^s=\mi n\) and
\(\sqrt g\,\mathfrak n^t=-\mi p\), we therefore obtain
\[
  \Xi_F^*\!\left(
    (\mathfrak n^\mu\lambda_\mu)\,\d\ell
  \right)
  =
  (\mi n\,\nu_s-\mi p\,\nu_t)\,\d\hat\ell.
\]
The same transformation applies to the moment flux.
Thus the metric factors cancel in these pulled-back flux--measure
products; in particular, no factor involving the arbitrary embedding
scale \(c_\tau\) remains.

\begin{remark}
The conservative completion is an algebraic consequence of the common co-normal flux structure.
It does not establish a discrete inf--sup bound for a chosen configuration--momentum pair.
The refinement study in Section~\ref{subsec:num_simplex_stability}
shows strong sensitivity to mesh resolution and the chosen
triangulation pattern for the investigated continuous equal-order
\(P_2/P_2\) simplex construction,
even though the formulation remains consistent under the stated
assumptions, the numerical flux is conservative, and the nonlinear
equations are solved to tight tolerance.
Accordingly, the simplex extension is retained as a diagnostic result rather than recommended as the default method.
\end{remark}

%------------------------------------------------------------
\subsection{Alternative momentum spaces}
\label{subsec:app_compatible_spaces}

The present computations use continuous equal-order spaces for the
configuration and independent momenta. A possible extension is to
reduce the temporal polynomial degree of the momenta. This choice is
motivated by the derivative couplings in the action form of
Subsection~\ref{subsec:worldsheet_mixed_form},
\[
  b_p(\mi p_h,\mi\eta_h)
  =\int_{\mathcal B_0}\mi p_h\cdot\mi\eta_{h,t}\,\d s\,\d t,
  \qquad
  b_\ell(\mi\ell_h,\mi\theta_h)
  =\int_{\mathcal B_0}\mi\ell_h\cdot\mi\theta_{h,t}\,\d s\,\d t.
\]
These terms pair the momenta with time derivatives of the configuration
tests. Let \(V_h\) and \(Q_h\) denote the underlying scalar spaces for
configuration tests and momenta, respectively, before initial and
test-trace conditions are imposed. A natural choice is to let the
momentum space contain these derivatives:
\begin{equation}
  \partial_tV_h\subset Q_h.
  \label{eq:app_derivative_compatible_spaces}
\end{equation}
For tensor-product splines on a common mesh, temporal differentiation
reduces degree \(p_t\ge1\) and continuity \(C^{\alpha_t}\), with
\(0\le\alpha_t\le p_t-1\), to degree \(p_t-1\) and continuity
\(C^{\alpha_t-1}\), while retaining the spatial basis.
For example, the derivative of a continuous piecewise quadratic
function is piecewise linear and may jump at temporal element
boundaries. The corresponding piecewise linear momentum space is
broken in time (\(C^{-1}\)): no continuity is imposed at these interfaces.

Using such a pair in the rearranged balance formulation of
Section~\ref{sec:spacetime_discretization} would require a consistent
treatment of initial data and momentum transfer across temporal
interfaces, together with a stability analysis of the resulting
discrete operator.

%------------------------------------------------------------------
% Bibliography
%------------------------------------------------------------------

\bibliographystyle{plain}
\bibliography{literature,literature_revision-8}

\end{document}